\documentclass[preprint,12pt]{elsarticle}

\usepackage{amssymb}
\journal{Computer Physics Communications}

\newcommand{\Ab}{{\bf A }}
\newcommand{\xb}{{\bf x }}

\usepackage{xcolor}

\usepackage{xspace}
\usepackage{color}
\usepackage{url}

\newcommand{\acro}{{\sc{SUperman}}\xspace}

\usepackage{algorithm, setspace}
\usepackage[noend]{algpseudocode}

\newcommand{\algrule}[1][.2pt]{\par\vskip.5\baselineskip\hrule height #1\par\vskip.5\baselineskip}
\usepackage{booktabs}
\usepackage{multirow}
\usepackage{array}
\usepackage{caption}
\usepackage{booktabs}
\usepackage{colortbl}
\usepackage{graphicx}
\usepackage{colortbl}
\usepackage{enumitem}
\usepackage{longtable}
\usepackage{subcaption}
\usepackage{amsmath}
\usepackage{placeins}

\usepackage{enumitem}
\newcommand{\revise}[1]{#1}

\begin{document}

\begin{frontmatter}



\title{{\acro}: Efficient Permanent Computation on GPUs}


\author[a]{Deniz Elbek}\ead{deniz.elbek@sabanciuniv.edu}
\author[a]{Fatih Taşyaran}\ead{fatihtasyaran@sabanciuniv.edu}
\author[b]{Bora U\c{c}ar}\ead{bora.ucar@ens-lyon.fr}
\author[a]{Kamer Kaya\corref{author}}

\cortext[author] {Corresponding author.\\\textit{E-mail address:} kaya@sabanciuniv.edu}
\address[a]{Sabanci University, Faculty of Engineering and Natural Sciences, Istanbul, T\"{u}rkiye}
\address[b]{CNRS, LIP (UMR5668 Universit\'e de Lyon - ENS de Lyon - UCBL - CNRS - Inria), Lyon, France; Institute for Data Engineering and Science (IDEaS), Georgia Institute of Technology, Atlanta, GA, USA}

\begin{abstract}
%A submitted program is expected to satisfy the following criteria: it must be of benefit to other physicists, or be an exemplar of good programming practice, or illustrate new or novel programming techniques which are of importance to computational physics community; it should be implemented in a language and executable on hardware that is widely available and well documented; it should meet accepted standards for scientific programming; it should be adequately documented and, where appropriate, supplied with a separate User Manual, which together with the manuscript should make clear the structure, functionality, installation, and operation of the program.

%Your manuscript and figure sources should be submitted through Editorial Manager (EM) by using the online submission tool at \\
%https://www.editorialmanager.com/comphy/.

%In addition to the manuscript you must supply: the program source code; a README file giving the names and a brief description of the files/directory structure that make up the package and clear instructions on the installation and execution of the program; sample input and output data for at least one comprehensive test run; and, where appropriate, a user manual.

%A compressed archive program file or files, containing these items, should be uploaded at the "Attach Files" stage of the EM submission.

%For files larger than 1Gb, if difficulties are encountered during upload the author should contact the Technical Editor at cpc.mendeley@gmail.com.
The {\em permanent} is a function, defined for a square matrix, with applications in various domains including quantum computing, statistical physics, complexity theory, combinatorics, and graph theory. Its formula is similar to that of the determinant; however, unlike the determinant, its exact computation is \#P-complete, i.e., there is no algorithm to compute the permanent in polynomial time unless P=NP.
For an $n \times n$ matrix, the fastest algorithm has a time complexity of $O(2^{n-1}n)$. Although supercomputers have been employed for permanent computation before, there is no work and, more importantly, no publicly available software that leverages cutting-edge High-Performance Computing accelerators such as GPUs. 
In this work, we design, develop, and investigate the performance of \acro, a complete software suite that can compute matrix permanents on multiple nodes/GPUs on a cluster while handling various matrix types, e.g., real/complex/binary and sparse/dense, etc., with a unique treatment for each type. 
\acro run on a single Nvidia A100 GPU is up to $86\times$ faster than a state-of-the-art parallel algorithm on 44 Intel Xeon cores running at 2.10GHz.
Leveraging 192 GPUs, \acro computes the permanent of a \revise{$62 \times 62$} matrix in 1.63 days, marking the largest reported permanent computation to date.

\noindent \textbf{PROGRAM SUMMARY}
\begin{small}
\noindent
{\em Program Title:}  \acro                                         \\
{\em CPC Library link to program files:} (to be added by Technical Editor) \\
{\em Developer's repository link:} {\url{https://github.com/SU-HPC/superman}}\\
{\em Licensing provisions(please choose one):} MIT \\
{\em Programming language:}  {\tt{C++}}, {\tt{CUDA}} \\
{\em Nature of problem:} 
The permanent plays a crucial role in various fields such as quantum computing, statistical physics, combinatorics, and graph theory. Unlike the determinant,  computing the permanent is \#P-complete [1] and its exact computation has exponential complexity. Even the fastest known algorithms require time that grows exponentially with matrix dimensions, making the problem computationally intractable for large matrices. The state-of-the-art tools leverage supercomputers [2, 3], but there remains a notable gap in publicly available software that exploits modern High-Performance Computing accelerators, such as GPUs. This limitation makes the researchers who require efficient and scalable methods for permanent computation suffer, particularly when dealing with various matrix types (real, complex, binary, sparse, dense) in practical applications. \\
{\em Solution method:} \acro is a complete open-source software suite built to tackle the computational challenges of permanent computation through a hybrid multi-node and multi-GPU approach. It supports a diverse range of matrix types, including real, complex, binary, sparse, and dense matrices, with specialized handling tailored to each type to maximize performance.  
In benchmarks, the software achieves an 86$\times$ speedup on a single GPU compared to state-of-the-art parallel algorithms running on 44 cores. Moreover, its scalable architecture allows for the distribution of workloads across multiple GPUs, enabling the computation of permanents for matrices as large as \revise{$62 \times 62$}—setting a new record in the literature. The modular design of \acro facilitates further research and development in high-performance permanent computation.\looseness=-1

%* Items marked with an asterisk are only required for new versions
%of programs previously published in the CPC Program Library.\\
\end{small}
   \end{abstract}
\end{frontmatter}


\section{Introduction}

For an $n \times n$ matrix $\mathbf{A}$, let $a_{i,j}$ be the entry at the $i$th row and $j$th column of $\mathbf{A}$. The permanent of $\mathbf{A}$ is defined as
\begin{equation}
{\tt perm}(\mathbf{A}) = \sum_{\sigma \in {\mathcal P}}{\prod_{i=1}^{n}a_{i, \sigma(i)}}\label{eq1}
\end{equation}
\noindent where $\mathcal P$ is the set of all permutations of the numbers  ${1,2, \dots, n }$, and $\sigma$ is a unique permutation. Although the formula is similar to that of the determinant, 
the permanent is \#P-complete~\cite{valiant79a} to compute.
This means that an exact computation of the permanent is infeasible for large enough matrices.
For an $n \times n$ matrix, the best algorithm for general matrices in the literature has a worst-case runtime complexity of $O(2^{n-1}n)$~\cite{ryser63,nijenhuis78}. 

Permanent is an important function used to understand the characteristics of a matrix and has various applications in physics, combinatorics, computer science, statistics and more. Matrix permanents emerge prominently in quantum physics, particularly in systems involving bosons. 
The role of permanents contrasts sharply with that of determinants in fermionic systems. 
For instance, in quantum optics, the probability amplitudes of photons traversing a linear optical network are governed by the permanent of a unitary matrix describing the network. 
This relates to boson sampling, a paradigm proposed by Aaronson and Arkhipov~\cite{aaronson11, LUNDOW2022110990}. 
In number theory, the sums of Fibonacci and Lucas numbers~\cite{kilic2007}, where the former is generated by the recurrence $F_0 = 0$, $F_1 = 1$, and $F_{n+2} = F_{n+1} + F_n$, for $n \geq 0$, and the latter is generated by $L_0 = 2$, $L_1 = 1$, and $L_{n+2} = L_{n+1} + L_n$, for $n \geq 0$, are directly related to the permanent of a special matrix family, i.e., lower-Hessenberg matrix family. 
The theory of permanents is also used in order statistics to synthesize results arising from the ordering of independent and non-identically distributed random variables~\cite{balakrishnan2007}. 
Furthermore, the permanent is used in DNA profiling in which the exact values are needed to find the exact genotype probability distributions~\cite{Narahara2013ApplicationOP}. 
For these applications, a tool that can compute the permanents of dense and weighted matrices is indispensable.

In the context of boson sampling, computing permanents of sparse matrices is interesting since low-depth boson sampling yields sparse matrices~\cite{Brod15}. 
Since graphs and sparse matrices have structural similarities, the permanents of sparse matrices correspond to combinatorial concepts on their corresponding graphs. In this context, permanents feature in statistical mechanics and combinatorial physics. For instance, they relate to counting problems, such as perfect matchings in bipartite graphs, which can be leveraged to model physical systems such as dimer arrangements on lattices~\cite{BEICHL1999128,PhysRevE.77.016706} due to the fact that the permanent of an $0$-$1$ adjacency matrix is equal to the number of perfect matchings~\cite{lovasz2009matching}. 
For directed graphs, the permanent of the $0$-$1$, binary adjacency matrix is equal to the number of vertex-disjoint cycle covers. This relation has also been used for a simple proof of \#P-completeness of permanent computation in~\cite{253457}. 
Such combinatorial quantities are essential while analyzing mathematical structures such as \textit{Nash equilibrium} in game theory \cite{Merschen11}. For these applications, a tool that can compute the permanents of sparse matrices is necessary. 

Recently, HPC clusters and supercomputers have been used in practice to compute permanents for boson sampling;
Wu~et~al.~\cite{10.1093/nsr/nwy079} report the time to compute the permanent of a $48 \times 48$ dense matrix on $8192$ nodes of the Tianhe-2 supercomputer ($196608$ cores) as $4500$ seconds. They also test an approach using both CPU cores and two MIC co-processors/node with 256 nodes to compute the permanent of a $40 \times 40$ matrix in 132 seconds. 
In another study, Lundow and Markstr\"{o}m used a cluster to compute the permanent of a record-breaking $54 \times 54$ matrix in 7103 core/days~\cite{LUNDOW2022110990}. For their experiments, the authors used 2 to 400 nodes of the Kebnekaise cluster with {\tt MPI} whose nodes are equipped with 14-core Intel Exon CPUs.\looseness=-1

In this work, we design, develop, and measure the performance of our open-source tool \acro, which can compute the matrix permanents using CPUs, GPUs, and High-Performance Computing~(HPC) clusters. \acro can handle both dense and sparse matrices, and matrices with real and complex entries. 
Our main motivation is to design and develop a tool for permanent computation that can be used in various domains, and on architectures equipped with multicore processors, multi-GPU machines or even on supercomputers. \revise{Although the tool can leverage multicore CPUs and multiple nodes, this paper mainly focuses on {\em{how to compute the permanent on a single manycore GPU efficiently}} since \acro leverages the techniques from the literature for the multi-node/GPU, and multi-core settings. 
The paper additionally examines the distributed memory case briefly, and reports experimental results on three different systems---one with a single Nvidia Quadro GV100, one having 8 nodes each having an Nvidia A100, and another with 48 nodes each having four Nvidia H200.} We make this tool publicly available\footnote{\url{https://github.com/SU-HPC/superman}} for reproducibility and usability.
Here are our contributions and some results on the performance of \acro:
\begin{itemize}
    \item Depending on the hardware at hand, \acro can utilize CPU cores via {\tt OpenMP}, GPUs via {\tt CUDA}, and multiple nodes via {\tt MPI}.
    \item Compared to a state-of-the-art parallel algorithm on a 44-core Intel Xeon Gold 6152 CPU, \acro is up to $59\times$ and $86\times$ faster on an Nvidia Quadro GV100 and an Nvidia A100, respectively. 
    \item Making use of multiple GPUs, 
    \acro can compute the permanent of a \revise{$62 \times 62$ matrix in 140989 seconds, i.e., 1.63 days, on 192 Nvidia H200 GPUs.} This matrix is larger than the previous record stated in~\cite{LUNDOW2022110990}. \revise{To be precise, the computation of the permanent of this matrix requires about $293.9\times$ more operations than the previous world record computation of the permanent of an $54\times 54$ matrix.}
\end{itemize}

The rest of the paper is organized as follows; Section~\ref{sec:back} introduces the background and notation followed in the paper. The GPU parallelization and optimizations used in \acro are summarized in Section~\ref{sec:algo:gpu}. Section~\ref{sec:acro:prep} describes the implementation details to obtain the desired precision level. The experimental results are given in Section~\ref{sec:exp}, and Section~\ref{sec:conc} concludes the paper and discusses further research avenues. 

\section{Background and Notation}\label{sec:back}
A naive algorithm based on the definition of the permanent~\eqref{eq1} enumerates all permutations $\sigma$ and computes $\prod_i a_{i,\sigma(i)}$ for each, yielding a worst-case runtime complexity of $\mathcal{O}(n \times n!)$.
Ryser~\cite{ryser63} leverages the {\em inclusion and exclusion} principle to rewrite the equation as 
\begin{equation}
{\tt perm}(\mathbf{A}) = (-1)^n  \sum_{S \subseteq \{1,2, \ldots, n\}} (-1)^{|S|} \prod_{i=1}^{n} \sum_{j \in S}{a_{i,j}}\;. \label{eq2}
\end{equation}
The worst-case complexity of evaluating the above formula is $\mathcal{O}(\sum_{k=1}^n n k \binom{n}{k})$. This is so, as for a set $S$ of $k$ columns, one performs $\mathcal{O}(n k)$ operations to compute $\prod_{i=1}^{n} \sum_{j \in S}{a_{i,j}}$ and there are $\binom{n}{k}$  sets of cardinality $k$.
As $\sum^n_{k=1} k\binom{n}{k} = n 2^{n-1}$, the worst-case 
runtime complexity is $\mathcal{O}(n^2 2^{n-1})$. 

Nijenhuis and Wilf improve Ryser's algorithm by processing the column sets $S$ in Gray code order~\cite[Ch. 23]{nijenhuis78}. 
Let {\sc{Gray}}$_{g}$ be the $g$th binary Gray code for a given number of bits; a 3-bit Gray code is shown in Table~\ref{tbl:3g}. For each sequence, the rightmost bit is the least significant one and has the id 0. 
The bit IDs corresponding to the changed bits with respect to the previous sequence are given in the 4th and 5th columns, respectively, for the natural order and Gray order. 
As the table shows, only a single bit changes between two consecutive Gray codes, whereas all three bits can change for the ordinary 3-bit binary code. 
For a Gray code, the $g$th code equals to 
$$\mbox{\sc{Gray}}_{g} = g \oplus (g \gg 1)$$ where $\oplus$ is the bitwise XOR operation and $g \gg 1$ shifts the bitwise representation of $g$ to the right by one bit.

\begin{table}
\begin{center}
\small
\begin{tabular}{l|rr|rr}
&\multicolumn{2}{c|}{Values}&\multicolumn{2}{c}{Changed Bits}\\\hline
&	{\sc{Binary}}	&{\sc{Gray}} &{\sc{Binary}}	&{\sc{Gray}}\\\hline
{\sc{Gray}}$_{0}$	&000	&000 & - & -\\
{\sc{Gray}}$_{1}$	&001	&001 & 0 & 0\\
{\sc{Gray}}$_{2}$	&010	&011 & 0,1 & 1\\
{\sc{Gray}}$_{3}$	&011	&010 & 0& 0\\
{\sc{Gray}}$_{4}$	&100	&110 & 0,1,2& 2\\
{\sc{Gray}}$_{5}$	&101	&111 & 0& 0\\
{\sc{Gray}}$_{6}$	&110	&101 &0,1 & 1\\
{\sc{Gray}}$_{7}$	&111	&100 & 0& 0\\
\end{tabular}
\end{center}
\caption{3-bit binary and Gray codes.}
\label{tbl:3g}
\end{table}
 
In Table~\ref{tbl:3g}, the set bits, which are $1$, in the {\sc{Gray}} column values correspond to the column sets $S$. For instance, {\sc{Gray}}$_{2} = 011$ corresponds to the set $S_2$ containing the first and the second columns of $\mathbf{A}$ (the counting starts from the least significant bit). With this ordering of the column sets, two consecutively processed sets $S_{g}$ and $S_{g+1}$ always differ by one bit/column as seen in the rightmost column of Table~\ref{tbl:3g}.  
Hence for $S_{g+1}$, processing the sets in Gray Code order makes the (row) summations to be used in the product $\prod_{i=1}^{n} \sum_{j \in S_{g+1}}{a_{i,j}}$ be ready in $\mathcal{O}(n)$ time, since these are done on top of $\sum_{j \in {S_g}}{a_{i,j}}$ which are already computed for $1 \leq i \leq n$. That is for each set, one needs $\mathcal{O}(n)$ operations to compute the product term $\prod_{i=1}^{n} \sum_{j \in S_{g+1}}{a_{i,j}}$ through $\prod_{i=1}^{n} \sum_{j \in S_{g}}{a_{i,j}}$. With this approach, the worst-case time complexity becomes 
 $\mathcal{O}(n2^{n})$. 

Another optimization by Nijenhuis and Wilf reduces the complexity to $\mathcal{O}(n2^{n-1})$~\cite{nijenhuis78}. Here, we reach the same optimization in a simpler way without the need to define permanent-like functions for rectangular matrices. Let 
$$\mathbf{A'}=\begin{pmatrix}
\mathbf{A}& \xb\\
\mathbf{0}_{1\times n} &1
\end{pmatrix}
$$
be an $(n+1)\times(n+1)$ matrix obtained by extending $\mathbf{A}$ with a row and column, where $\mathbf{0}_{1\times n}$ is a row vector of size $n$ with all zero elements, and  $\xb=[x_1, x_2, \ldots, x_n]^T$.  The $x_i$s will be specified later. 
In the last row of  $\mathbf{A'}$, the entries below the  
 columns of $\mathbf{A}$ are all zeros. Since in the same row the $(n+1)$st entry is one, we have ${\tt perm}(\mathbf{A'}) = {\tt perm}(\mathbf{A})$ by the definition of the permanent in~\eqref{eq1}. Writing the inclusion and exclusion formula~\eqref{eq2} for $\mathbf{A'}$ yields:
$${\tt perm}(\mathbf{A'}) = {\tt perm}(\mathbf{A}) = (-1)^{n+1}  \sum_{S' \subseteq \{1,2, \ldots, n+1\}} (-1)^{|S'|} \prod_{i=1}^{n+1} \sum_{j \in S'}{a'_{i,j}}\;.$$
Let us split the summation into two, where the first one runs over subsets $S'$ that contain $n+1$, and the second one runs over the others:
\begin{align}
{\tt perm}(\mathbf{A}) = &\left((-1)^{n+1}  \sum_{\stackrel{S' = \{n+1\} \cup S}{S\ \subseteq \{1,2, \ldots, n\}}} (-1)^{|S'|} \prod_{i=1}^{n+1} \sum_{j \in S'}{a'_{i,j}}\right) 
+ \\&\left((-1)^{n+1}  \sum_{{S'\ \subseteq \{1,2, \ldots, n\}}} (-1)^{|S'|} \prod_{i=1}^{n+1} \sum_{j \in S'}{a'_{i,j}}\right).\label{eq4}
\end{align}
\noindent The second summation evaluates to zero, since the last term in the product, ${\sum_{j\in S'}} a_{n+1, j}$, is zero for all $S'$ not containing the $(n+1)$th column.

In the first summation, we can take the product term for $i=n+1$ out, as
 $\sum_{j\in S'}a'_{i,j}=1$, and write
$${\tt perm}(\mathbf{A}) = \left((-1)^{n+1}  \sum_{\stackrel{S' = \{n+1\} \cup S}{S\ \subseteq \{1,2, \ldots, n\}}} (-1)^{|S'|+2} \prod_{i=1}^{n} \sum_{j \in S'}{a'_{i,j}}\right)\;.$$
As $|S'|=|S|+1$,  $(-1)^{|S|+3} = (-1)^{|S|-1}$, we can manipulate $(-1)^{n+1}$ and $(-1)^{|S'|+2}$
and 
explicitly write the contribution of the $(n+1)$st column in each row to obtain
\begin{equation}{\tt perm}(\mathbf{A}) = (-1)^{n}  \sum_{{S\ \subseteq \{1,2, \ldots, n\}}} (-1)^{|S|} \prod_{i=1}^{n} \left(x_i + \sum_{j \in S}{a_{i,j}}\right)\;.\label{eq:nwx}\end{equation}
Let $S$ be a subset of $\{1,2, \ldots, n\}$ and $\overline{S} =  \{1,2, \ldots, n\} \setminus S$ be its complement. The product term for $\overline S$ in the  equation~\eqref{eq:nwx} can be written using that for $S$ as
$$\prod_{i=1}^{n} \left(x_i + \sum_{j \in \overline{S}}{a_{i,j}}\right) =  \prod_{i=1}^{n} \left(x_i + \sum_{j=1}^n a_{i,j} - \sum_{j \in {S}}{a_{i,j}}\right)\;.$$
Setting $x_i = -\frac{1}{2}\sum_{j=1}^n a_{i,j}$ yields, respectively, the following product terms for $S$ and its complement $\overline{S}$ in~\eqref{eq:nwx}
$$ \prod_{i=1}^{n} \left(\sum_{j \in S}{a_{i,j}} -\frac{1}{2}\sum_{j=1}^n a_{i,j}\right) \mbox{ and } \prod_{i=1}^{n} \left(\frac{1}{2}\sum_{j=1}^n a_{i,j} - \sum_{j \in S}{a_{i,j}}\right)\;.$$
As the product term for $\overline{S}$ is equal to $(-1)^n$ times that for $S$, computing only one of them is enough for~\eqref{eq:nwx}.  
That is, one can process $2^{n-1}$ subsets $S\subseteq\{1,2,\ldots,n\}$ to compute the permanent of $\mathbf{A}$.

Nijenhuis and Wilf take one more action to be able to use Gray Code as well to reduce the worst-case runtime complexity to $\mathcal{O}(n2^{n-1})$. They divide the sets $S\subseteq\{1,2,\ldots,n\}$ into two groups. In each group, no two sets complement each other. The complements of the sets in the first group are therefore in the second group and vice versa.
They achieve this by putting all $S\subseteq\{1,2,\ldots,n-1\}$ in the first group, and  $\overline{S}=S\cup\{n\}$ in the second group, which is the complement of a set in the first group.
This way, the sets $S\subseteq\{1,2,\ldots,n-1\}$ can be created with the help of Gray Code, 
and the stated runtime complexity can be achieved.
For all the implementations in {\acro}, we have used this variant. The base implementation is given in Algorithm~\ref{alg:ryser}. 

\begin{algorithm}[H]
\footnotesize

\caption{: {\sc Ryser} }\label{alg:ryser}
\textbf{Input:} $\Ab$ \algorithmiccomment{$n \times n$ matrix} \\
\textbf{Output:} {\tt perm}($\Ab$)\algorithmiccomment{Permanent of  $\Ab$}

\algrule
\begin{algorithmic}[1]
\Statex
\For{$i = 1$ to $n$}  
	\State{$sum = \sum_{j = 1}^n a_{i,j}$} \algorithmiccomment{For each row of $\Ab$, compute the row sum}
	\State{$\xb[i] \leftarrow a_{i,n} - \frac{sum}{2}$ } \algorithmiccomment{Apply Nijenhuis \& Wilf modification}
\EndFor
 \algrule

\State{$p \leftarrow \prod_{i = 1}^n{\xb[i]}$} \algorithmiccomment{Start with an empty column subset of $\{1, \ldots, n-1\}$ }

 \algrule

 \For {$g = 1$ to $2^{n-1} - 1$} \label{ln:loop1} \algorithmiccomment{For all subsets $S \subset \{1, \ldots, n-1\}$}
	\State{$j \leftarrow \log_2$({\sc{Gray}}$_{g}$ $\oplus$ {\sc{Gray}}$_{g-1}$)}\label{ln:j} 
    \State{$s \leftarrow 2 \times ${\sc{Gray}}$_{g}$[$j$] $-1$}\algorithmiccomment{If $j$ is being added $s = 1$, o.w., $s= - 1$}
	 \State{$prod \leftarrow 1$} \algorithmiccomment{The contribution for the column set at hand}
	 \For {$i =  1$ to $n$} \label{ln:loop2} 
	        \State{$\xb[i] \leftarrow \xb[i] + (s \times a_{i, j})$}\label{ln:xb} \algorithmiccomment{Update row sums based on $s$ }
	        \State{$prod \leftarrow prod \times \xb[i]$}	        
	 \EndFor

	 \State{$p \leftarrow p + \left((-1)^g \times prod\right)$}\label{ln:p} \algorithmiccomment{Add the product to the permanent}
\EndFor 
	  \algrule

\State{\Return{$p \times (4 \times (n \bmod 2) - 2)$}} \algorithmiccomment{Double and multiply with $(-1)^{n-1}$}
 \end{algorithmic}
\end{algorithm}

Parallel permanent computation for dense matrices has been investigated in various studies. 
For large matrices, this is expensive, and parallelization is extremely useful. Lundow and Markström~\cite{LUNDOW2022110990} present parallelization of the permanent computations on CPU-based, shared-memory, and distributed-memory systems. 
Wu~et~al.~\cite{10.1093/nsr/nwy079} 
computed the permanents of matrices with $n > 40$ on tens of thousands of nodes and hundreds of thousands of cores of a supercomputer. \revise{In both studies, as well as in this work, the parallelisation is performed on the loop at line~\ref{ln:loop1} where the iteration space is distributed to nodes/cores.}
Lundow and Markström report $10^{-9.2} \times n \times 2^n$ core-seconds for their implementation, which implies 19.4 years of computation time on a single core for $n = 54$. 
\acro, on the other hand, can compute the permanent of a \revise{$62 \times 62$ matrix in 1.63 days via 192 H200 GPUs.}

\revise{
The base algorithm for \acro proposed in this work is similar to the one applied by Lundow and Markström~\cite{LUNDOW2022110990}, shown also in Algorithm~\ref{alg:ryser}. On the other hand, Wu~et~al.~\cite{10.1093/nsr/nwy079} used a recomputation-based approach and computed $prod$ in Algorithm~\ref{alg:ryser} from scratch without keeping an auxiliary $\xb$ array. This is why their Ryser variant has complexity $\mathcal{O}$($n^22^n$) instead of $\mathcal{O}$($n2^{n-1}$). Our preliminary experiments revealed that for a $40 \times 40$ matrix, this variant can be $25\times$ slower than Algorithm~\ref{alg:ryser} with Nijenhuis and Wilf optimisation. In addition, Wu~et~al. experimented with Balasubramanian-Bax-Franklin-Glynn~(BB/FG) formula which also yields $\mathcal{O}$($n^2 2^n$) complexity. 

Although Algorithm~\ref{alg:ryser} is fast thanks to the $\xb$ array being maintained and efficiently updated in Gray-code order, it makes the computation memory-intensive. For $n = 40$, a naive GPU implementation that stores the array in shared memory spends $70\%$ of its time on memory accesses. Hence, a better utilisation of the GPU memory hierarchy is required.
One of the main contributions of this work is how to do this efficiently, as explained in Section~\ref{sec:algo:gpu}.}

\subsection{Permanent Computation for Sparse Matrices}

There are studies in the literature which exploit sparsity for computing sparse-matrix permanents~\cite{kaya:19, forbert03, mittal01, liang06}. For a sparse matrix, the number of matrix transversals (a set of $n$ nonzeros in an $n \times n$ matrix, each coming from different rows/columns) can be small. Since every nonzero product in Eq.~\ref{eq1} corresponds to a transversal and every transversal yields a nonzero product, the idea of enumerating all the matrix transversals is promising and has been investigated by Mittal and Al-Kurdi~\cite{mittal01}.
Doing this while exploiting the computational and storage efficiency requires the exploitation of sparse data structures, Compressed Row Storage~(CRS) and Compressed Column Storage~(CCS). As Figure~\ref{fig:crs} shows, an $m \times n$ sparse matrix $\mathbf{A}$ with $nnz$ nonzeros can be stored in CRS and/or CCS formats, which can be considered dual to each other. The CRS can be defined as follows:

\begin{figure}[htbp]
    \centering
    \includegraphics[width=.9\textwidth]{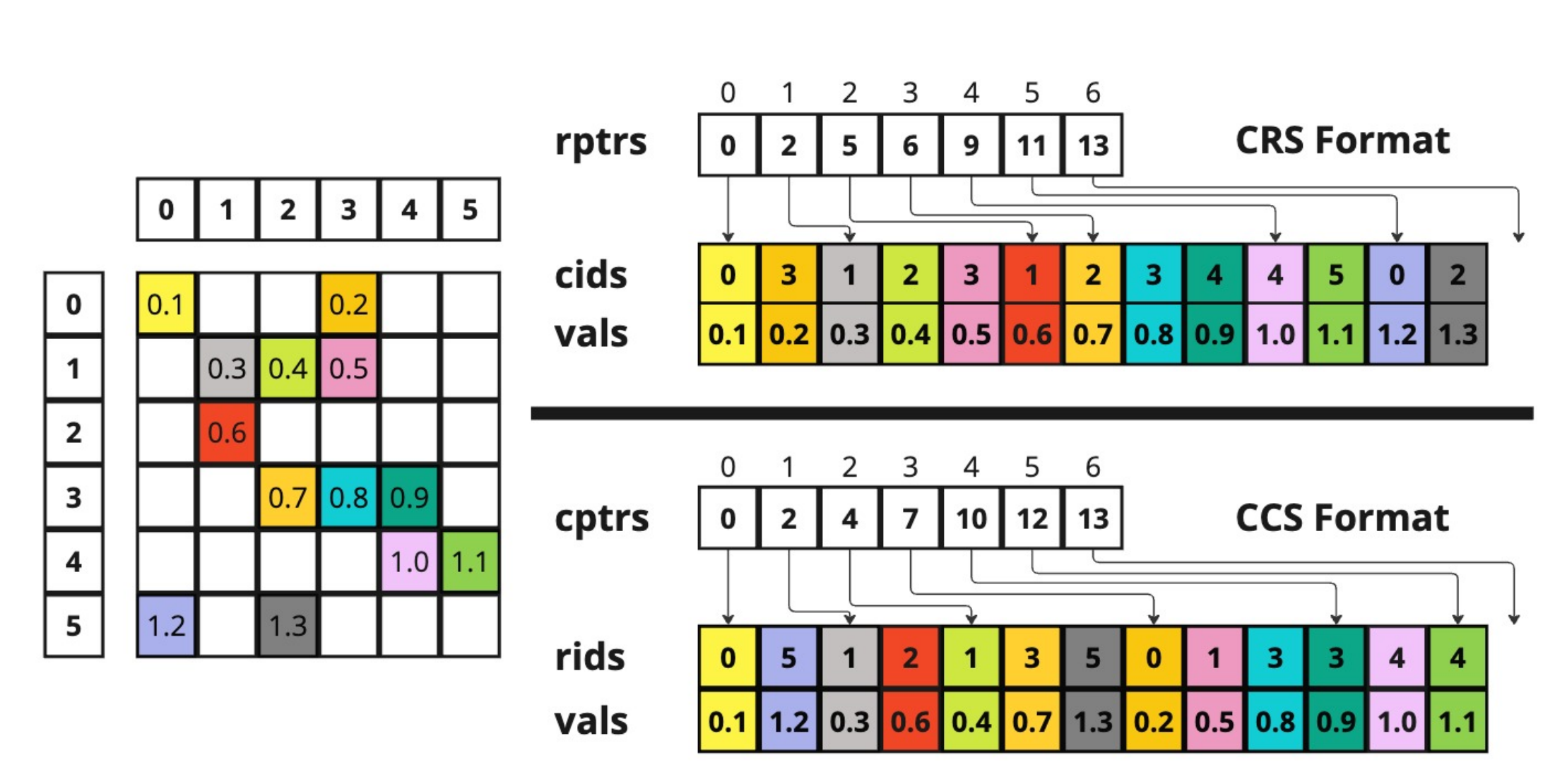}
    \caption{CRS and CCS formats for a $6 \times 6$ matrix with 13 nonzeros. In CRS/CCS, the first element of {\tt rptrs}/{\tt cptrs} arrays is 0, and their last element is equal to 13. In CRS, the {\tt cids} and {\tt vals} arrays store the nonzeros in row-major order, whereas in CCS, the {\tt rids} and {\tt vals} arrays store them in column-major order.}\label{fig:crs}
\end{figure}

\begin{itemize}
    \item {\tt rptrs[$\cdot$]} is an integer array of size $m + 1$. For $0 \leq i < m$, {\tt rptrs[i]} is the location of the first nonzero of the $i$th row in the {\tt cids} array. The first element is {\tt rptrs[0]} $= 0$, and the last element is {\tt rptrs[m]} $ = nnz$. Hence, all the column indices of row $i$ are stored between {\tt cids[rptrs[i]]} and {\tt cids[rptrs[i + 1]] - 1}. 
    \item {\tt cids[$\cdot$]} is the integer array storing the column IDs for each nonzero in row-major ordering. 
    \item Each nonzero value $val$ is stored in an array {\tt cval} in the same order of nonzeros inside {\tt cids}.
\end{itemize}

Enumerating all the transversals can be efficient for extremely sparse matrices having only a few transversals. On the contrary, when the number of transversals is large, which is usually the case for real-life sparse matrices, the cost of enumeration increases significantly. 
Hence, a direct adaptation of Algorithm~\ref{alg:ryser} for sparse matrices using CRS and CCS, which is shown in Algorithm~\ref{alg:sparyser}, is the main algorithm used in the literature~\cite{kaya:19}.

\begin{algorithm}[htbp]
\footnotesize
\caption{: {\sc SpaRyser}} 
\label{alg:sparyser}
\textbf{Input:} ($rptrs,  cids, rvals$) \algorithmiccomment{CRS of $n \times n$ sparse matrix $\Ab$} \\ \hspace*{8.2ex}($cptrs, rids, cvals$) \algorithmiccomment{CCS of $n \times n$ sparse matrix $\Ab$}\\
\textbf{Output:} {\tt perm}($\Ab$) \algorithmiccomment{Permanent of $\Ab$}

\algrule
\begin{algorithmic}[1]
\Statex
\For{$i = 1$ to $n$} 
	\State{$sum = 0$}  \algorithmiccomment{Row sum for the $i$th row}
	\For{$ptr = rptrs[i]$ to ($rptrs[i+1] - 1$)}  
		\State{$sum \leftarrow sum + rvals[ptr]$}
	\EndFor
	\State{$\xb[i] \leftarrow a_{i, n} - \frac{sum}{2}$} \algorithmiccomment{Alg. 1: N/W variant}	 
\EndFor

 \algrule
	\State{$p \leftarrow \prod_{i = 1}^n{\xb[i]}$}

 \algrule
 \For {$g = 1$ to $2^{n-1} - 1$} \label{ln:loop1spa} \algorithmiccomment{For all $S \subset \{1, \ldots, n-1\}$}
	\State{$j \leftarrow \log_2$({\sc{Gray}}$_{g}$ $\oplus$ {\sc{Gray}}$_{g-1}$)}\label{ln2:j}
	\State{$s \leftarrow 2 \times ${\sc{Gray}}$_{g}$[$j$] $-1$}	 
	  \For {$ptr = cptrs[j]$ to ($cptrs[j + 1] - 1$)} \algorithmiccomment{Nnzs of the $j$th column}
	        \State{$row \leftarrow rids[ptr]$}
	        \State{$val \leftarrow cvals[ptr]$}

	 \EndFor

	 	\State{$prod \leftarrow 1$}  \algorithmiccomment{The contribution for this Gray code}	 
	 	\For {$i =  1$ to $n$} 
	        		\State{$prod \leftarrow prod \times \xb[i]$}      
	 	\EndFor
		\State{$p \leftarrow p + \left((-1)^g \times prod\right)$}
\EndFor 

\algrule
\State{\Return{$p \times (4 \times (n \bmod 2) - 2)$}} \algorithmiccomment{Double and multiply with $(-1)^{n-1}$}
 \end{algorithmic}
\end{algorithm}

\subsection{Managing Storage on GPUs for Permanent Computation}
For the GPUs, there are various places where the data can be stored. Table~\ref{tab:gpu_memory_hierarchy} presents the memory hierarchy. 
For permanent computation, the shared data is the matrix $\Ab$ and is small. The private data is the row-sum array $\xb$ of size $n$ per thread. Although the data is small, the number of arithmetic operations is exponential in $n$ and each operation requires access to $\Ab$ and $\xb$. 
As usual for most applications, how the data are stored has the utmost impact on the memory access efficiency and the overall performance. CPUs and GPUs leverage different computation patterns and architectural optimizations, and the optimal layout changes w.r.t. the architecture. 

\begin{table}[H]
\small
    \centering
    \scalebox{0.8}{
    \begin{tabular}{ p{1.5cm} | p{7.6cm} | p{2.6cm} | p{1.5cm}  }
    \textbf{Mem. Type} & \textbf{Description} & \textbf{Scope} & \textbf{Latency} \\
    \hline
    Registers & 
    The fastest memory, used by each thread for local variables and intermediate calculations. Each thread has its own set of registers. & 
    Private to each thread. & 
    Very low \\
    \hline
    Shared Memory & 
    Memory shared among threads in the same block. Faster than global memory but slower than registers. Shared memory can be explicitly managed by programmers. & 
    Shared within a thread block. & 
    Low  \\
    \hline
    Constant Memory & 
    Read-only memory used to store constant values for all threads. Access is cached, making it faster than global memory for small amounts of data. & 
    Read-only and shared across the GPU. & 
    Moderate \\
    \hline
    Global Memory & 
    The largest but slowest type of memory. Accessible by all threads across all blocks. & 
    Accessible by all threads. & 
    High  \\
    \hline
    Local Memory & 
    Memory allocated for each thread's private use but stored in global memory. Used when registers are insufficient. & 
    Private to each thread. & 
    High  \\
    \end{tabular}
    }
    \caption{GPU Memory Hierarchy}
    \label{tab:gpu_memory_hierarchy}
\end{table}

\section{\acro: Computing Permanents on GPUs}\label{sec:algo:gpu}

The computation of the original permanent equation~\eqref{eq1} or that of the modified one~\eqref{eq2} are both pleasingly parallelizable, since each term inside the leftmost summation is independent of each other. 
However, when Gray code is leveraged, each iteration depends on the previous one. 
That is, the $\xb$ vector in Algorithm~\ref{alg:ryser} is not computed from scratch but updated, and this renders the iterations interdependent. 
With $\tau$ threads, it is possible to divide the summation into $\tau$ chunks with size $chnksz = \lceil{\frac{2^{n-1} - 1}{\tau}}\rceil$, assign these chunks to the threads, and initialize private $\xb$ vectors for each chunk separately based on the initial Gray value. 
This idea, borrowed from the literature~\cite{kaya:19}, is shown in Algorithm~\ref{alg:parryser}.  
With this approach, the threads can {\em independently} compute partial permanent values whose sum equals the actual value $p_{global}$~(line~\ref{ln:pglobal} of Alg.~\ref{alg:parryser}). 
Note that the same parallelization strategy can be applied to CPUs and GPUs, as well as to sparse matrices.

\begin{algorithm}[H]
\footnotesize
\caption{: {\sc ParRyser}} \label{alg:parryser}

\textbf{Input:} $\Ab$ \algorithmiccomment{$n \times n$ matrix} \\
 \textbf{Output:} {\tt perm}($\Ab$)\algorithmiccomment{Permanent of  $\Ab$ }

\begin{algorithmic}[1]
 \algrule
 \Statex
\State {$\cdots$ Lines 1-4 of Algorithm~\ref{alg:ryser} $\cdots$}
\algrule
 \setcounter{ALG@line}{4}
\For{$tid = 0$ to $\tau - 1$ {\bf in parallel}} \algorithmiccomment{$\tau$ is the number of threads}
 \State{$\xb_{tid} \leftarrow \xb$}  \algorithmiccomment{Copy $\xb$ to the private copy}
  \State{$chnksz \leftarrow \lceil{\frac{2^{n-1} - 1}{\tau}}\rceil$} \label{alg:chunksize} \algorithmiccomment{No of iterations per thread, e.g., chunk size}
 \State{$start \leftarrow (chnksz * tid) + 1$}  \algorithmiccomment{First iteration for $tid$}
  \State{$end \leftarrow \min(2^{n-1}, start + chnksz - 1)$} \algorithmiccomment{Stopping iteration for $tid$}

 \hrulefill
  \State{$\triangleright$ \footnotesize Set the state of column sums in $\xb_{tid}$ to the previous iteration}
 \For{$j = 1$ to $n$} 
        \If{{\sc{Gray}}$_{start - 1}[j] = 1$} \algorithmiccomment{If the $j$th bit of previous code is 1}
    	\For{$i = 1$ to $n$} \algorithmiccomment{Add the $j$th column to $\xb_{tid}$} 
    		\State{$\xb_{tid}[i] \leftarrow \xb_{tid}[i] + a_{i,j}$}
    	\EndFor
        \EndIf
\EndFor
 \algrule

 \For {$g = start$ to $end$} \label{ln:looppara1}
	\State ... \\ \hspace*{7.1ex}Lines 6--11 of Algorithm~\ref{alg:ryser}\\\hspace*{7.1ex}...
        \setcounter{ALG@line}{21}
        \State{$p_{global} \leftarrow p_{global} \stackrel{{atomic}}{+}  \left((-1)^g \times prod\right)$}\label{ln:pglobal} \algorithmiccomment{Update the global var.}
\EndFor
\EndFor
\algrule
\State{\Return{$p_{global} \times (4 \times (n \bmod 2) - 2)$}} \algorithmiccomment{Double and mult. with $(-1)^{n-1}$}
\end{algorithmic}
\end{algorithm}

\subsection{Storing the Matrix on GPU}
\begin{figure}[htbp]
    \centering
    \includegraphics[width=.9\textwidth]{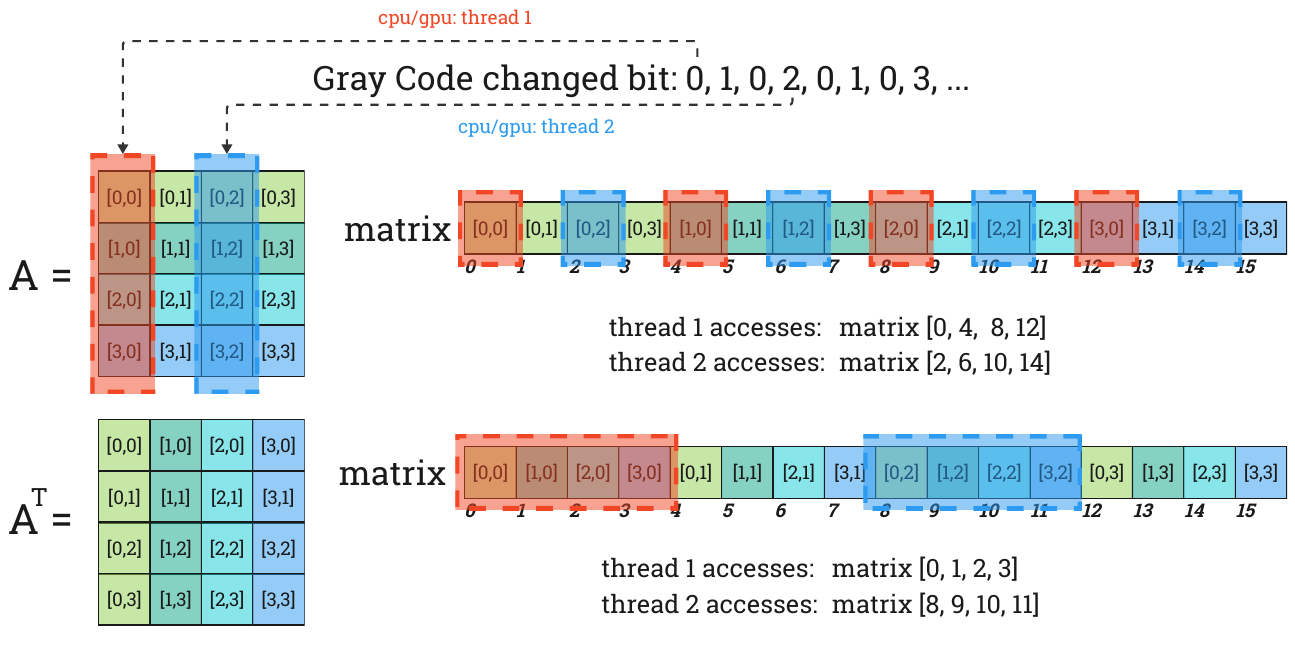}
    \caption{Row-major~(top) and column-major~(bottom) layout storage when the matrix $\Ab$ is kept in global memory. {\sc Ryser} requires access to matrix columns, which implies strided access to the matrix elements for row-major storage. This causes many cache misses for CPUs. On the other hand, with the column-major layout, once a CPU thread touches a location, it fetches the adjacent content in the same cache line to its cache. However, the column-major layout is problematic for GPUs. GPU threads act as teams of 32 and always execute their (memory) operations at the same time. When the threads access far-away locations, it causes stalls due to expensive memory access. The impact can be mitigated when the matrix is stored in row-major format.}
    \label{fig:memory-layout}
\end{figure}

As seen before, except for the first few steps initializing $\xb$, all algorithms presented in this paper access the matrix by columns to update $\xb$ in $2^{n-1}$ iterations. 
This is why for CPUs, storing $\Ab$ in the column-major layout is the best option, both for the dense and the sparse case, since this allows the column entries to be cached after the first access~(see the bottom part of Figure~\ref{fig:memory-layout}). 
On the contrary, if the matrix is stored in {\em global memory} of a GPU, the column-major layout is not the best option, due to the GPUs' {\em Single Instruction Multiple Threads}~(SIMT) paradigm. With SIMT, the memory is accessed concurrently by a group of threads within the same instruction. 
This group is called a {\em warp}, and the number of threads in a single warp is 32 for modern GPUs. 
For a data load/store instruction, we define a {\em coalesced access} as the case where all memory accesses are spatially adjacent in the 1D global/device memory layout, residing within the same 128B memory block.
On the contrary, if the accesses are spread to multiple blocks, the accesses are called {\em uncoalesced} and will be more costly. 
Consider an iteration (line~\ref{ln:looppara1} of Algorithm~\ref{alg:parryser}), in which the threads in a warp process different column IDs, i.e., the changed bit indexes for are different within their Gray codes $g$. 
In this case, if the matrix \Ab is stored in the column-major layout, the accessed locations, i.e., $a_{i,j}$ entries at line~\ref{ln:xb} of Algorithm~\ref{alg:ryser}, spread more compared to the row-major layout. Hence, if the matrix is stored in global memory it may be better to use row-major layout.

A {\em streaming multiprocessor} (\revise{SM}) can be considered as the main, independent processing unit of a GPU. 
A GPU contains multiple \revise{SM}s, and each \revise{SM} unit consists of multiple cores capable of executing hundreds of threads simultaneously. 
An \revise{SM} possesses a limited amount of {\em shared memory}, a programmable on-chip memory that allows threads within a block to communicate and exchange data efficiently. 
Unlike global memory, which is accessible by all threads but relatively slow, shared memory is much faster because it resides on-chip. 
Although the amount of shared memory per \revise{SM} is limited, around $48$KB--$164$KB per \revise{SM}, it is usually sufficient to store the matrices for which permanent will be computed. 
For instance, an $60 \times 60$ dense matrix can be stored in $28.8$KB even in double precision (note that the record for permanent computation until this work is $n = 54$). 

When the global memory is used, $\Ab$ is stored only once. 
On the contrary, when the shared memory is used, it is stored on each \revise{SM}~(8--108 \revise{SM} exist depending on the GPU architecture), and multiple times on each \revise{SM}, i.e., once for each {\em thread block}, a group of warps scheduled together on the same \revise{SM}. 
This may yield a problem and reduce GPU utilization due to the limited shared memory size and the inability to run the maximum number of threads on a single \revise{SM}. 
For instance, for a GPU with $48$KB shared memory per \revise{SM}, if $\Ab$ requires more than $24$KB memory to be stored, only one block can reside on an \revise{SM} which can support up to 2048 threads, whereas a block can contain 1024 threads. 
Hence, half of the capacity will not be used.  
Fortunately, modern GPUs contain more than $48$KB~(e.g., 64KB--96KB) and for permanent computation, two blocks, containing 1024 threads each, can concurrently run on a single \revise{SM}.

\begin{figure}[h]
    \centering
    \includegraphics[width=0.93\textwidth]{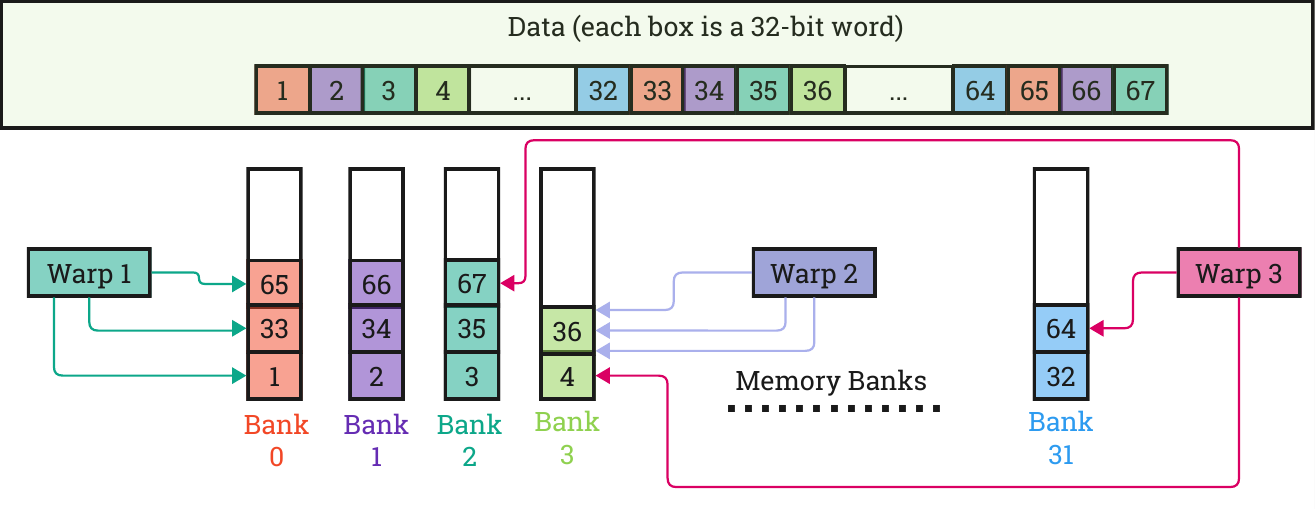}
    \caption{67 consecutive 32-bit words in GPU shared memory are distributed to the 32 memory banks in a round-robin fashion. The threads inside three warps perform memory access: the first warp requests three different elements from Bank 0, hence, a 3-way bank conflict occurs. All the threads in the second warp request the same item from Bank 3. This is called a broadcast; there is no bank conflict since no serialisation occurs. All of the third warp's accesses are to different banks, and there is no bank conflict for these accesses.}
    \label{fig:bank_conflict}
\end{figure}

\subsubsection{Coalescing Memory Accesses and Eliminating Bank Conflicts}\label{sec:algo:ceg}

To enable high-throughput memory accesses, shared memory is divided into 
32 {\em memory banks} and the data items 
of size 32 bits
are distributed to these banks in a round-robin fashion. 
For larger items, e.g., numbers in double precision,  adjacent banks are used for the same data item. 
Each bank can be accessed independently, enabling multiple threads to read or write simultaneously without {\em conflicts}. 
However, if two or more threads in the same warp attempt to access the same memory bank (for different data items), a {\em bank conflict} occurs, resulting in the serialization of those accesses, which can slow down performance. 
If all accesses from a warp are to the same data item, a broadcast operation is performed, and no conflicts are generated. 
Figure~\ref{fig:bank_conflict} shows how a toy data block is stored in memory banks with three warps accessing these banks. 
The first warp creates a bank conflict, whereas the second and third do not.   

When the matrix $\Ab$ is stored in column-major format, and $n$, the number of rows/columns, is a multiple of 32, the $i$th entries of each column reside on the same bank. 
Hence, when different bits are changed in the Gray codes, i.e., when the values at line~\ref{ln:j} of Alg.~\ref{alg:ryser}~(or line~\ref{ln2:j} of Alg.~\ref{alg:sparyser}) are different for the threads in the same warp, each column access creates a bank conflict (as in Warp 1 in Figure~\ref{fig:bank_conflict}). 
Although there will be fewer of them, bank conflicts will still occur since, regardless of the value $n$, columns [0, 32, 64, $\ldots$] will have the same entry alignment over the memory banks. 
They will have their $i$th entries in the same bank for $0 \leq i < n$. 
However, they will hurt performance more when $n$ and 32 are not coprime. 
For instance, when $n = 40$, i.e., when $gcd(n, 32) = 8$, the alignments of columns $j = 4k + \ell$ for the same $0 \leq \ell < 4$ over the memory banks will be the same. 
That is by setting $\ell = 0$, one can see that when accessed at the same time, columns $[0, 4, 8, \ldots, 32, 36]$ will incur bank conflicts. 
Hence, if the 0th and 4th bits of the Gray codes change for two different threads in the same warp within their corresponding local iterations there will be one. 
In general, bank conflicts can be avoided in two different ways; one can either have an access pattern as that of Warp 2 or Warp 3 in Fig.~\ref{fig:bank_conflict}. 
We will follow the former, i.e., make the changed Gray code bits the same within the same local iteration of all the threads in a warp as much as possible. 

\begin{figure}[H]
    \centering
     \begin{subfigure}[b]{\textwidth}
        \includegraphics[width=\textwidth]{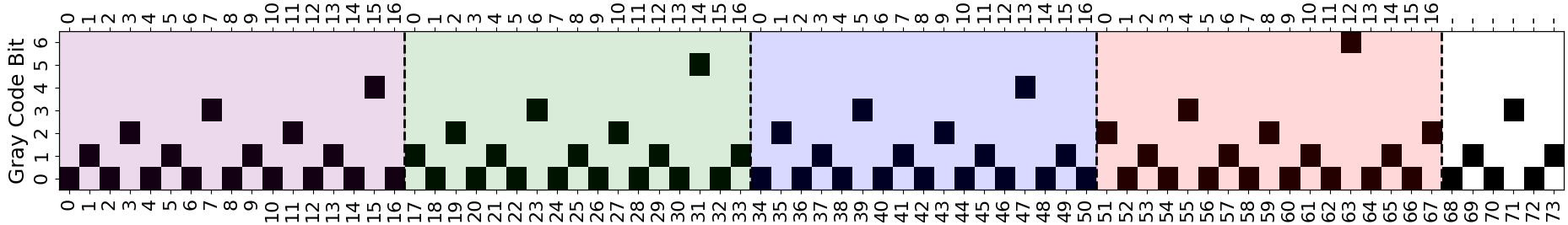}
        \caption{The changed Gray code bits of iterations, sorted w.r.t. the global iteration IDs. The chunk size is $17$.}
        \label{fig:bc11}
     \end{subfigure}
    \centering
    \begin{subfigure}[b]{\textwidth}
        \includegraphics[width=\textwidth]{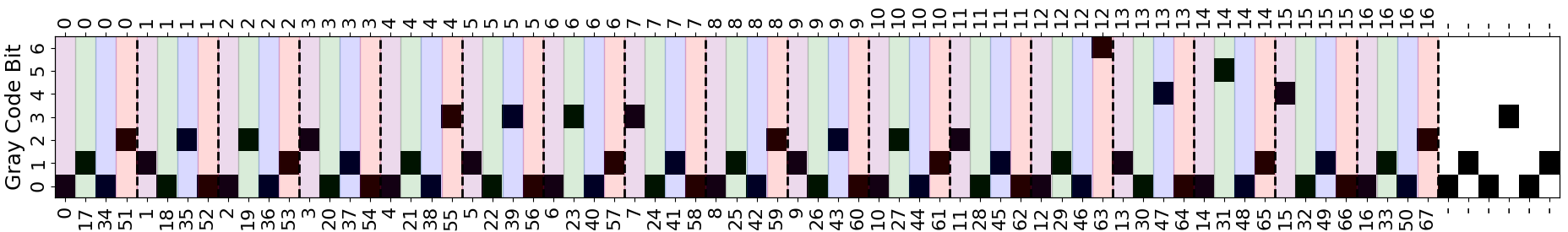}
        \caption{The changed Gray code bits of iterations, sorted w.r.t. the local iteration IDs which is the actual execution order with SIMT. As the figure shows, each set of 4 columns with the same local iteration ID yields memory accesses to 3 different columns.}
        \label{fig:bc12}
     \end{subfigure}
     \caption{The changed bits ($y$-axis) of 4 threads with a chunk size equal to 17. Different colours represent different threads. The numbers above and below are local and global iteration IDs. }
    \label{fig:normal}
\end{figure}

To yield the maximum performance from a GPU kernel, i.e., a function to be executed on the device, selecting the launch parameters correctly is crucial. 
The choice of launch parameters affects the occupancy, which is the ratio of active warps (groups of threads) to the maximum number of warps supported on a multiprocessor. 
Because each GPU is different in its capabilities, such as the number of SMs, shared memory available per SM, number of {\tt CUDA} cores, etc., it is clear that selecting hard-coded parameters empirically will drastically affect performance negatively when run on a different GPU. 
However, {\tt CUDA} run-time has helper functions to select the correct kernel launch parameters for such cases.
Originally, we used the {\tt CUDA API}  to program the device. 
Given the kernel, i.e., a device code, {\tt CUDA} suggests the optimal number of threads, $\tau$ to maximize device occupancy. 
Then the chunk size, the number of iterations assigned to each thread, is set to $\lceil{\frac{2^{n-1} - 1}{\tau}}\rceil$ as in line~\ref{alg:chunksize} in Algorithm~\ref{alg:parryser}. 
When the chunk size is chosen arbitrarily, the changed bits can often be different for threads in a warp. Figure~\ref{fig:normal} illustrates this for 4 threads inside a warp with a chunk size equal to 17. 
At each local iteration performed by this set of threads, 3 different columns are accessed.

\begin{figure}[H]
    \centering
     \begin{subfigure}[b]{\textwidth}
        \includegraphics[width=\textwidth]{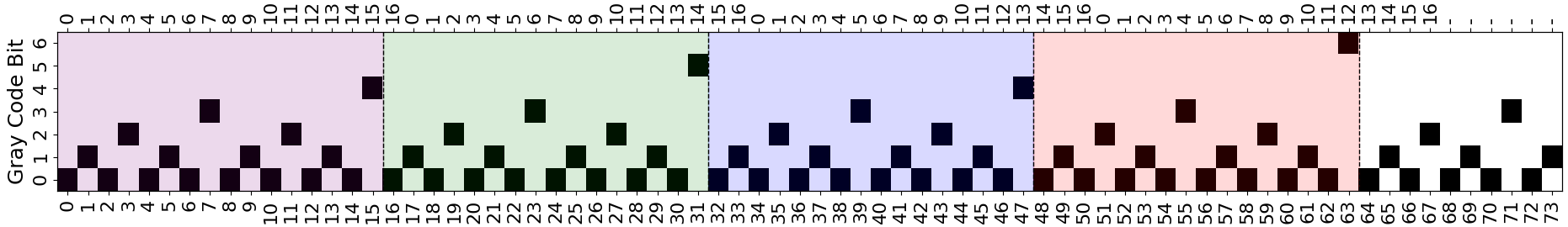}
        \caption{The changed Gray code bits of iterations, sorted w.r.t. the global iteration IDs. The chunk size is $16$.}
        \label{fig:bc21}
     \end{subfigure}
    \centering
    \begin{subfigure}[b]{\textwidth}
        \includegraphics[width=\textwidth]{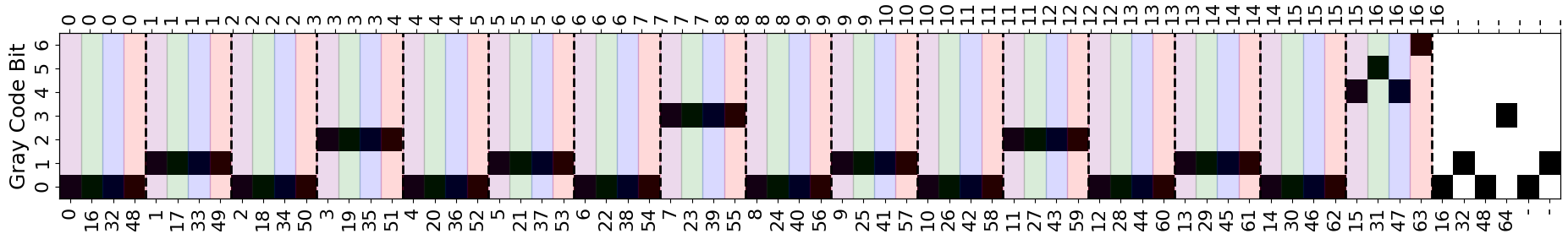}
        \caption{The changed Gray code bits of iterations, sorted w.r.t. the local iteration IDs, which is the actual execution order via SIMT. As the figure shows, each set of 4 columns with the same local iteration ID yields memory accesses to the same column except for the last local iterations.}
        \label{fig:bc22}
     \end{subfigure}
     \caption{The changed bits ($y$-axis) of 4 threads with a chunk size equal to 16. Different colours represent different threads. The numbers above and below are local and global iteration IDs.}
    \label{fig:nice}
\end{figure}

The $n$-bit Gray code sequence can be constructed recursively from the $(n-1)$-bit Gray code by first reflecting the existing sequence (i.e., reversing the order of the elements) and concatenating it to itself. Then the 1-bit prefix {\color{blue}{0}} is used for the original sequence, and {\color{red}{1}} is used for the reversed~\cite{knuth}. For instance, the $3$--bit code $\{{\color{blue}{0}}00, {\color{blue}{0}}01, {\color{blue}{0}}11, {\color{blue}{0}}10 | {\color{red}{1}}10, {\color{red}{1}}11, {\color{red}{1}}01, {\color{red}{1}}00\}$ is obtained from the $2$--bit code $\{{\color{blue}{0}}0, {\color{blue}{0}}1 |  {\color{red}{1}}1, {\color{red}{1}}0\}$. This makes the sequence of changed bit locations for an $(n-1)$-bit Gray code recursively appear twice within that of the $n$-bit Gray code. Let $CBL_k$ be the changed bit locations for a $k$--bit code.
For instance, for a $3$--bit code is the string $CBL_3 = [0,1,0,2,0,1,0]$, which is a palindrome, i.e., the same as its reverse. 
The changed bit locations for the $4$-bit code are then $CBL_4 = CBL_3 + [3] + {CBL^R_3}$ where ${CBL^R_3} = {CBL_3}$ denotes the reverse of ${CBL_3}$, ``[3]'' makes a string, and the binary operator $+$ concatenates the left and right strings. 
With induction, it holds that $$CBL_n = CBL_{n-1} + [n-1] + {CBL^R_{n-1}}$$ for every $n$. Note that the length of $CBL_n$ is $2^n-1$. 
Following these, it can be observed that if the chunk size is a power of $2$, then the sequences of the changed bit locations for the threads will be the same except for the last local iteration. 
This is illustrated for a toy case in Figure~\ref{fig:nice}. 
To realize an almost bank-conflict-free execution, we get $\tau$ from the API as in the original implementation, then compute the chunk size and round it down to the closest power of 2. 
Once this kernel is executed, the remaining set of iterations is handled with another kernel, which is almost the same as the initial execution but much cheaper. 
 
As a last note, making the threads work on the same column is not useful only for eliminating bank conflicts on shared memory. It is also useful for coalescing accesses to global memory since the threads within a warp will access the same address during their local iterations for most of the time. Hence, the proposed modification is expected to increase the performance regardless of the placement of the matrix $\Ab$. 

\subsection{Storing the $\xb$ Array on GPU with Matrix Specific Rebuilds}\label{sec:algo:rebuild}

Since $\xb$ is thread-private, in \acro, we store it on the thread's local memory. The {\tt CUDA} runtime allocates local memory for each thread upon kernel launch. Access to this region is expensive, as it is logically located in the same region as the global memory. 
However, the data in this region is frequently referenced throughout the kernel execution. 
This contradiction leads the compiler to perform an important optimization called \textit{register allocation}, during which the data located in the thread-local memory is cached into thread-private registers. This allows threads to have fast access to the data they use most frequently. 
However, as stated by the {\tt CUDA} Programming Guide~\cite{NVIDIA}, the register(s) to be accessed should be discoverable at compile time. 
The most trivial approach to enable this is to allocate a local memory for $\xb$ by assuming a maximum size $n_{max}$, known during the compilation. During the permanent computation, only $n \leq n_{max}$ of these locations are used.  
In \acro, we have used $n_{max} = 63$. However, this may incur inefficiencies, since the registers are scarce. 

An optimization we leverage is rebuilding \acro by providing the dimension, $n$, of the matrix to the build process. That is $n_{max} = n$ is used to eliminate inefficiency. Although this makes the executable specific to $n$ and not usable for another matrix dimension $n' \neq n$, it allows the compiler to perform optimizations as stated before. The cost of compilation is negligible since the complexity of permanent computation is exponential with respect to $n$ whereas the compilation takes less than 3 seconds. This is why in \acro, $n_{max}$ can be set to $n$ by the user to rebuild the executable for the matrix at hand. Such optimization is not essential for {\em shared} or {\em global} memory whose sizes can be set dynamically during runtime.

For sparse matrices, $n$ is not the only value that has an impact on the memory access pattern. Since \acro employs a sparse CRS/CCS representation, the nonzero indices inside {\bf cptrs} and {\bf rids} also change the addresses to be accessed throughout the execution. Hence, only using $n_{max} = n$ and leaving the rest to the compiler is not sufficient to store $\xb$ on registers. This is why, given a sparse matrix, \acro automatically generates a kernel for each column which contains the row indices as constant literals. With these kernels, the compiler exactly knows which registers will be accessed, and register allocation optimization becomes possible. For further details on this approach, we refer the reader to~\cite{elbek2025fullyautomatedcodegenerationefficient}.

\subsection{Computing Permanent on Multiple GPUs and Multiple Nodes}

\revise{Regarding the parallelization strategy on multiple GPUs within a single node, our library supports both OpenMP- and MPI-based approaches, with the choice determined by the user through the configuration file. Specifically, when the configuration is set with \texttt{"mode" = "multi\_gpu"}, \texttt{"gpu\_num"} equals to the number of GPUs in the node, and \texttt{"processor\_num" = 1}, the computation is parallelized within the node using OpenMP, provided that at least \texttt{gpu\_num} threads are available for the process. 

When the configuration is set with \texttt{"mode" = "multi\_gpu\_mpi"}, \texttt{"gpu\_num" = 1}, and \texttt{"processor\_num"} equals to the number of GPUs in the node, \acro launches \texttt{processor\_num} tasks, each bound to a single GPU, and the computation is parallelized within the node through MPI. 
With this flexibility, one can even mix the parallelization strategies by setting \texttt{"mode" = "multi\_gpu\_mpi"}, \texttt{"gpu\_num"} to the number of GPUs in each node, and \texttt{"processor\_num"} to the number of nodes to enable inter-node parallelization with MPI and intra-node parallelization with OpenMP. An example distribution to six GPUs residing on three nodes is illustrated in Figure~\ref{fig:hierarchical_parallelization}.

\begin{figure}[htbp]
    \centering
    \includegraphics[width=\linewidth]{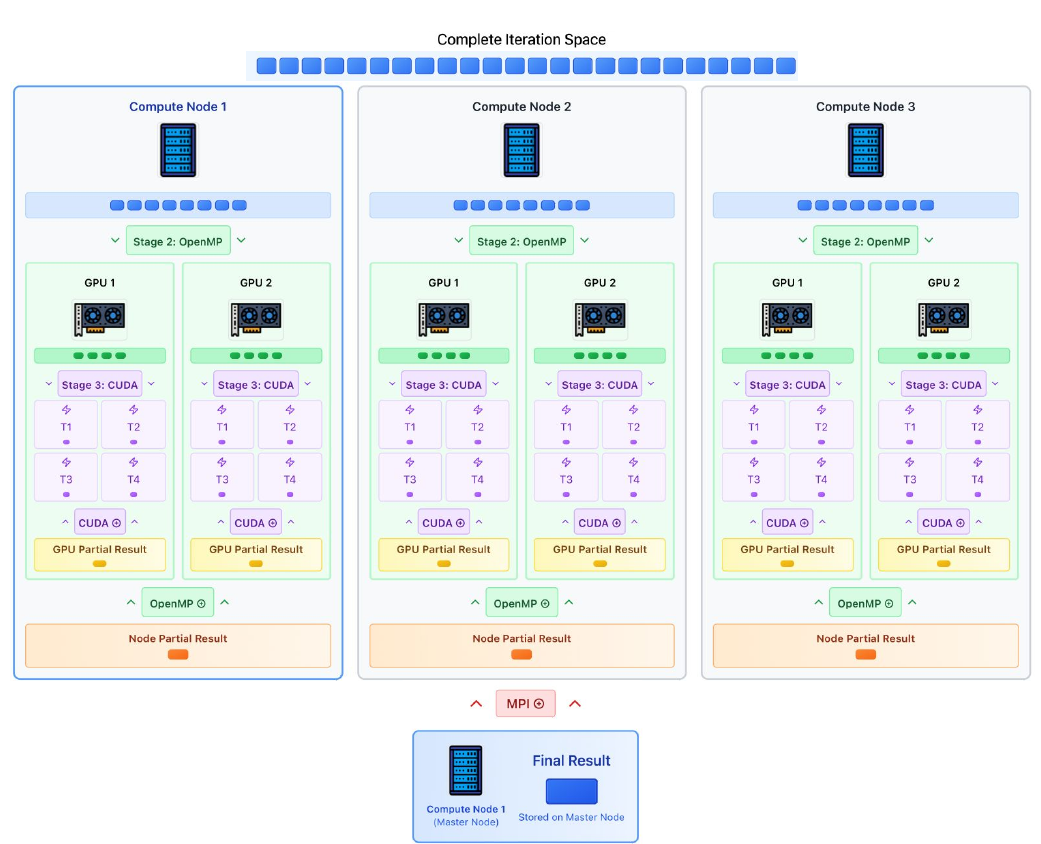}
    \caption{\revise{Hierarchical parallelization in \acro with MPI to 3 nodes and OpenMP to 2 GPUs at each node: The iteration space is divided into 3 parts by using MPI, which are sent to 3 nodes. Each part is then divided into 2 inside a single node via OpenMP where one thread is managing one GPU.}}
    \label{fig:hierarchical_parallelization}
\end{figure}

For the $62 \times 62$ matrix, we used the naive full MPI approach, where the parallelization is done entirely with MPI, in which \texttt{"processor\_num"} is set to the number of GPUs in the entire cluster. This setting does not differentiate between GPUs located within the same DGX node and those distributed across multiple nodes, both of which are supported in \acro.}

\section{\acro's Preprocessing Sparse Matrices}\label{sec:acro:prep}
Extra improvements to exploit the sparsity of the matrices have been investigated for permanent computation. For \acro, (1) we propose using the Dulmage-Mendelsohn~(DM) decomposition~\cite{Dulmage_Mendelsohn_1958} to remove nonzeros which do not have an impact on the value of the permanent, and (2) we adopt a decomposition technique from the literature~\cite{forbert03}. To the best of our knowledge, the former technique, using DM to eliminate redundant entries, has not been used in the literature for exact permanent computation. 

\subsection{Preprocessing: Elimination of Redundant Entries\label{sec:DMDecompose}}
 Let $G_{\Ab} = (V_R \cup V_C, E)$ be an undirected bipartite graph obtained from the matrix $\mathbf{A}$, whereas $V_R$ and $V_C$ are disjoint vertex sets corresponding to rows and columns of $\mathbf{A}$, respectively. 
 The edge set $E$ contains an edge $\{r_i, c_j\}$ for $r_i \in V_R$ and $c_j \in V_C$ if and only if there is a nonzero entry $a_{i,j}$ in $\mathbf{A}$. 
 For an $n \times n$, square matrix $\mathbf{A}$, a {\em perfect matching} $\mathcal{M}$ of $G_A$ contains $n$ edges 
 from $E$ such that each vertex in $V_R$ and $V_C$ is an endpoint of exactly one edge in $\mathcal{M}$.

 For a more efficient computation of a sparse matrix permanent, as a preprocessing step, we eliminate the nonzeros in $\mathbf{A}$ that do not contribute to the permanent. This is accomplished by making use of the {\em fine-grain Dulmage-Mendelsohn}~(DM) decomposition~\cite{Dulmage_Mendelsohn_1958,pofa:90} which partitions the vertices of $G_{\Ab}$ into subsets such that two vertices belong to the same subset if and only if they are matched in at least one perfect matching of $G_{\Ab}$. The decomposition process starts with a perfect matching $\mathcal{M}$, which can be found in 
 $\mathcal{O}(m \sqrt{n})$ time for an $n \times n$ sparse matrix with $m$ nonzeros~\cite{doi:10.1137/0202019}. Let $G'_{\Ab} = (V_R \cup V_C, E')$ be the {\em directed} bipartite graph obtained from $G_{\Ab}$ by orienting the edges in $E$ in a way that all edges in $\mathcal{M}$ are directed from row vertices, $V_R$, to column vertices, $V_C$, and all other edges in $E \setminus \mathcal{M}$ are oriented in the reverse direction. 
 
\begin{figure}[htbp]
    \centering
    \includegraphics[width=.8\textwidth]{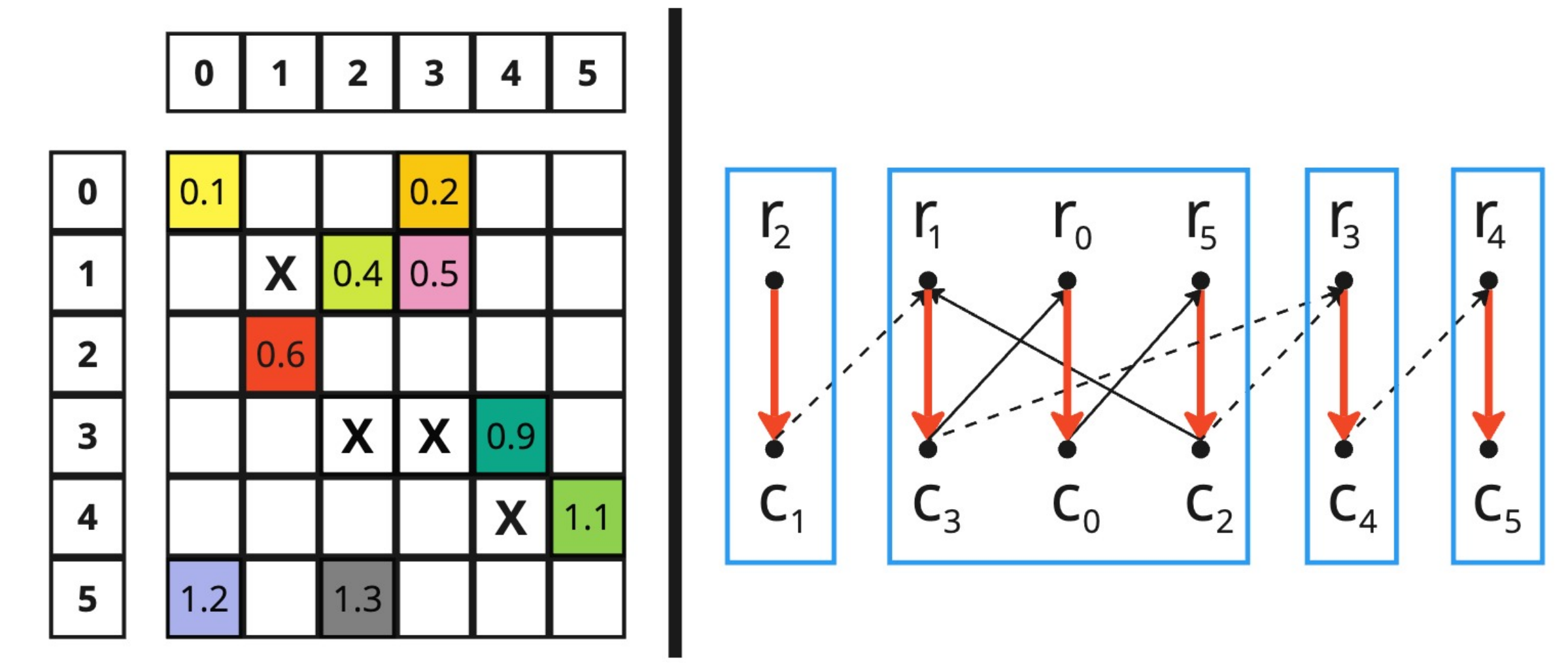}
    \caption{The fine-grain Dulmage Mendelsohn decomposition~(on the right) for the toy matrix $\Ab$ presented in Fig.~\ref{fig:crs}. There exist 4 SCCs for the directed bipartite graph obtained from $\Ab$ and 4 edges (dashed) whose endpoints are not contained in a single component. The corresponding 4 entries are removed from $\Ab$~(on the left) by marking them with {\bf{X}}.}\label{fig:dm}
\end{figure}

 Figure~\ref{fig:dm}~(right) shows the directed bipartite graph $G'_{\Ab}$ for the toy matrix presented in Fig.~\ref{fig:crs}. 
 The DM decomposition generates the {\em strongly connected components}~(SCCs) of $G'_{\Ab}$. A graph is said to be {\em strongly connected} if each vertex is reachable from every other vertex. The SCCs of $G'_{\Ab}$, which are the blue rectangles in Figure~\ref{fig:dm}~(right), are the maximal, strongly connected subgraphs of $G'_{\Ab}$. 
 These components can be identified in $\mathcal{O}(m + n)$ time~\cite{doi:10.1137/0201010}. 
 
 Based on the property of DM decomposition, an edge $\{r_i, c_j\} \in E$, corresponding to an entry $a_{i,j}$ of $\Ab$, is included in none of the perfect matchings if $r_i$ and $c_j$ belong to different SCCs of $G'_{\Ab}$. 
 There is a one-to-one correspondence between a perfect matching and a permutation $\sigma$ in Eq.~\eqref{eq1} with a {\em nonzero contribution} to the permanent; for a given $\sigma$, the corresponding edge set is $\{r_1, c_{\sigma(1)}\}, \{r_1, c_{\sigma(2)}\}, 
 \ldots, \{r_1, c_{\sigma(n)}\}$. 
 The contribution $\prod_{i=1}^{n}a_{i, \sigma(i)}$ is nonzero if all these edges are in $E$, i.e., they form a perfect matching. Hence, all the nonzeros whose corresponding edges are not contained in a single SCC can be deleted. 
 This way of discarding edges based on the DM decomposition is known and used earlier for approximately counting perfect matchings~\cite{dkpu:22}. 
 In Figure~\ref{fig:dm}~(left), we have four such entries marked with $\mathbf{X}$, each of which corresponds to an edge between different SCCs shown in Figure~\ref{fig:dm}~(left) with a dashes line.
 These nonzero entries can be deleted because they cannot be included in a permutation with a nonzero contribution to the permanent. 
 
 Following its steps, the DM decomposition is expected to be useful when there exist many SCCs in $G'_{\Ab}$. For instance, in a lower-triangular matrix with all zero entries above the main diagonal, the decomposition removes all the off-diagonal entries. % and only leaves $n$ of them.  
 
\subsection{Decomposing $\mathbf{A}$ into Smaller Matrices\label{sec:FMDecompose}}

Forbert and Marx~\cite{forbert03} presented a matrix decomposition scheme for permanent computations based on the following equation:
\begin{align}
{\tt perm}(\Ab) &= 
{\tt perm}
\left(
\begin{bmatrix}
    \alpha    & \beta& \mathbf{c} \\
  \mathbf{d}      & \mathbf{e} & \mathbf{B} 
\end{bmatrix}
\right)\nonumber
\\
&=
{\tt perm}
\left(\begin{bmatrix}
     0   & 0& \mathbf{c} \\
  \mathbf{d}      & \mathbf{e} & \mathbf{B}
  \end{bmatrix}
\right)
+
{\tt perm}
\left(
\begin{bmatrix}
    \alpha\mathbf{e}  + \beta\mathbf{d}  & \mathbf{B} 
    \end{bmatrix}
\right)
\label{eq:expand}
\end{align}
where the input $\begin{bmatrix}
     \alpha   & \beta& \mathbf{c} \\
  \mathbf{d}      & \mathbf{e} & \mathbf{B}
  \end{bmatrix}
$ is an $n \times n$ matrix with $\alpha$ and $\beta$ are nonzero scalars, $\mathbf{c}$ is an ($n-2$)-dimensional row
vector, $\mathbf{d}$ and $\mathbf{e}$ are ($n-1$)-dimensional column vectors, and $\mathbf{B}$ is the remaining ($n - 1$) $\times$ ($n - 2$) submatrix.
Previous studies~\cite{kaya:19,liang06} 
show that this decomposition can reduce the runtime.

After performing the Dulmage-Mendelsohn decomposition and erasing the redundant entries, we adopt the decomposition from Forbert and Marx to \acro in a way that the sparse matrices are decomposed until all the rows/columns in each matrix generated during this process contain more than four nonzeros. 
Algorithm~\ref{algo:decompose} describes how \acro adopts this decomposition.

Let $minNnz$ be the minimum number of nonzeros in a row/column of $\Ab$. 
When $minNnz \leq 2$, only one of the generated matrices will have a nonzero permanent since in Eq.~\ref{eq:expand}, $\mathbf{c}$ will be an all $0$ vector. Hence, $${\tt perm}\left(\begin{bmatrix} 
     0   & 0& \mathbf{c} \\
  \mathbf{d}      & \mathbf{e} & \mathbf{B}
  \end{bmatrix}\right) = 0.$$ 
For these cases, {\sc{d1compress}}~(line \ref{ln:d1}) and {\sc{d2compress}}~(line \ref{ln:d2}) used in Algorithm~\ref{algo:decompose} output a single matrix $\Ab' = \begin{bmatrix}
    \alpha\mathbf{e}  + \beta\mathbf{d}  & \mathbf{B} 
    \end{bmatrix}$. Furthermore, when $\beta$ is $0$ and hence $minNnz = 1$, $\Ab'$ is set to $\begin{bmatrix}
    \mathbf{e}  & \mathbf{B} 
    \end{bmatrix}$ to obtain a better precision, and the returned value is multiplied by $\alpha$ since ${\tt perm}
\left(
\begin{bmatrix}
    \alpha\mathbf{e}  + \beta\mathbf{d}  & \mathbf{B} 
    \end{bmatrix}
\right) = \alpha \times {\tt perm}
\left(
\begin{bmatrix}
    \mathbf{e}  & \mathbf{B} 
    \end{bmatrix}
\right)$.   
When  $2 < minNnz \leq 4$, we continue to apply the decomposition process~(line \ref{ln:d34} of Algorithm~\ref{algo:decompose}). In this case, two matrices are generated as described in~\eqref{eq:expand}. The first matrix $\Ab'$ is an $n \times n$ matrix,
whereas the second $\Ab''$ is  $(n-1) \times (n-1)$. 
Our preliminary experiments show that applying the decomposition for larger values of $minNnz$ increases the execution time of the overall permanent computation. 

\acro also combines the decomposition scheme given in Algorithm~\ref{algo:decompose} with the appropriate permanent computation algorithm. We observed that when the density of the nonzeros, $nnz / n^2$, is at least $30\%$, the parallel dense algorithm  {\sc ParRyser} is faster. On the other hand, when sparsity is more than $70\%$, {\sc ParSpaRyser} is applied, which can be considered as the parallel version of Algorithm~\ref{alg:sparyser} with matrix-specific compilation. 

\begin{algorithm}[htbp]
\footnotesize
\caption{: \sc{DecompRyser}}
\label{algo:decompose}
\small 
\textbf{Input:} \textbf{A}: an $n \times n$ sparse matrix \\ 
\textbf{Output:} {\tt perm}($\Ab$) \algorithmiccomment{Permanent of \textbf{A}, calculated recursively}
\algrule

\begin{algorithmic}[1]
  \State{$\triangleright$ \footnotesize Below, $rc$ is a row (or column) having $minNnz$ nonzeros}
    \State $(rc, minNnz) \leftarrow ${\sc{getMinimumNnz}}$(\mathbf{A})$ 
    \algrule
    \If{$minNnz = 1$}
 
    \State $\textbf{A}' \leftarrow$ \Call{d1compress}{\textbf{A}} \label{ln:d1}
    \algorithmiccomment{{\small{Remove $rc$ and its matching row/column}}}
    \State \Return $\alpha$ $\times$ \Call{DecompRyser}{$\textbf{A}'$} \algorithmiccomment{\small{$\textbf{A}'$ is ($n-1$) $\times$ ($n-1$)}}
    \EndIf
    
    \If{$minNnz = 2$}    
    \State $\textbf{A}' \leftarrow$ \Call{d2compress}
    {\textbf{A}}      \algorithmiccomment{{\small{Remove $rc$}: $\Ab' = \begin{bmatrix}
    \alpha\mathbf{e}  + \beta\mathbf{d}  & \mathbf{B} 
    \end{bmatrix}$ from~\eqref{eq:expand}}}
    \State \Return \Call{DecompRyser}{$\textbf{A}'$}\label{ln:d2}\algorithmiccomment{ $\textbf{A}'$ is ($n-1$) $\times$ ($n-1$)}
    \EndIf
    
    \If {{$minNnz = 3$} {\bf {or}} {$minNnz = 4$}}
        \State $\textbf{A}', \textbf{A}''$ $\leftarrow$ \Call{d34compress}{\textbf{A}}\label{ln:d34}        \algorithmiccomment{\small $\textbf{A}'$ is $n \times n$ and $\textbf{A}''$ is $(n-1) \times (n-1)$}

        \State \Return \Call{DecompRyser}{\textbf{A}$'$} + \Call{DecompRyser}{\textbf{A}$''$}
    \EndIf
    \algrule

\If {the sparsity is over 70\%} \algorithmiccomment{Compute the permanent}
\State{\Return {\sc ParSpaRyser}($\Ab$)} 
\Else 
\State{\Return {\sc ParRyser}($\Ab$)} 
\EndIf
\end{algorithmic}
\end{algorithm}

\section{Increasing the Accuracy of \acro }\label{sec:acro:prec}

In theory, the addition and multiplication of real numbers are associative; however, in practice, they are not. This discrepancy arises because real numbers are represented in finite precision on digital machines, typically via floating-point representations of the form $\pm (1 + m) \cdot 2^e,$
where \(e\) denotes the exponent and \(m\) is the fractional part of the normalized mantissa. 
Real numbers that cannot be represented exactly in this form are rounded to the nearest representable value, with the rounding performed by the processor's floating-point unit before the value is stored in memory. Goldberg \cite{goldberg} showed that the rounding error for a single floating-point operation---such as addition or multiplication---can be as much as \(0.5\) \textit{ulp} (units in the last place), corresponding to a relative error of \(2^{-53}\) in the IEEE 754 double-precision standard (the standard on which almost all modern computers are based). This bound is expressed as:

\begin{equation}\label{eq:relative-error}
\text{Relative Error: } \frac{\lvert \text{Computed} - \text{Exact} \rvert}{\text{Exact}} \le 2^{-53} \approx 1.11\times10^{-16}\,.
\end{equation}

\noindent When performing a large number of floating-point operations, these individual rounding errors can accumulate and lead to a significant deviation from the exact result. For example, to compute the permanent of a \(45 \times 45\) matrix using Algorithm~\ref{alg:ryser}, one must perform approximately \(1.62\times10^{15}\) floating-point operations, potentially accumulating a maximum error of about \(18\%\). 
The simplest solution to mitigate this is to use higher-precision (e.g., quad-precision) data types, although higher precision significantly increases computation time. 
An alternative is to employ compensated summation---such as Kahan \cite{kahan} summation---which maintains an additional compensation term to correct for accumulated rounding errors.

In \acro, we modify our kernels to incorporate these techniques for improved accuracy. 
As previous studies~\cite{LUNDOW2022110990}, to test the effectiveness of the proposed techniques, we execute our kernels on matrices whose permanents are precisely known. 
Such matrices of size \(n\) can be generated with all entries equal to \(a\), yielding a permanent of \(n! \times a^n\). 
In all calculations, we use quad precision for the outer-sum variable (the variable to which the threads' partial results are reduced). 
This approach is found to be highly effective for improving precision and has little impact on execution time, as the outer sum is updated only as many times as there are threads available on the kernel. 
For the other variables---namely, the inner-product variable~(corresponding to $prod$ in Alg.~\ref{alg:ryser}) and the thread-private, partial result variable~(corresponding to $p$ in Alg.~\ref{alg:ryser}), we varied their precision and reported these results in Table~\ref{tab:prec}. 
The first four columns show the variations in data types used: the first letter denotes the type of the inner-product variable, and the second letter denotes the type of the partial result variable. 
For instance, \textbf{DQ} indicates that the inner-product is stored in {\bf{D}}ouble precision while the thread-partial result is in {\bf{Q}}uad precision. 
Because {\tt CUDA} does not natively support quad precision, we implemented the basic arithmetic operations using two doubles from scratch. 
The Kahan summation~\cite{kahan}, on the other hand, is used solely for the safe accumulation of the inner-product variable into the thread-partial result, corresponding to the {\em atomic add} operation in line 22 of Algorithm~\ref{alg:parryser}. 
For each implementation, we reported the execution time and the relative error, computed as in the left-hand side of Eq.~\eqref{eq:relative-error}.

\begin{table}
\center
\scalebox{0.80} {
\begin{tabular}{c|r|r|r|r|r}
\multicolumn{6}{c}{\textbf{Time}} \\
\cline{2-6}
\(n\) & \textbf{DD} & \textbf{DQ \cite{quad_article}} & \textbf{DQ \cite{quad_forum}} & \textbf{QQ} & \textbf{Kahan \cite{kahan}} \\
\hline
35\(\times\)35 & 1.31 & 1.36 & 1.47 & 2.63 & 1.34 \\
40\(\times\)40 & 14.25 & 15.60 & 16.34 & 60.12 & 14.78 \\
45\(\times\)45 & 496.34 & 557.99 & 592.17 & 2178.91 & 526.46 \\
48\(\times\)48 & 4118.14 & 4540.18 & 4716.39 & 17881.54 & 4430.63 \\
50\(\times\)50 & 17154.49 & 18395.20 & 20270.01 & Timeout & 17692.18 \\
\multicolumn{6}{c}{} \\
\multicolumn{6}{c}{\textbf{Relative Error}} \\
\cline{2-6}
\(n\) & \textbf{DD} & \textbf{DQ \cite{quad_article}} & \textbf{DQ \cite{quad_forum}} & \textbf{QQ} & \textbf{Kahan \cite{kahan}} \\
\hline
35\(\times\)35 & 1.40e-09 & 9.70e-12 & 1.06e-11 & 9.70e-12 & 8.78e-12 \\
40\(\times\)40 & 2.69e-07 & 6.51e-11 & 8.56e-11 & 6.51e-11 & 6.63e-11 \\
45\(\times\)45 & 1.49e-04 & 2.54e-10 & 2.31e-10 & 2.54e-10 & 2.34e-10 \\
48\(\times\)48 & 1.09e-05 & 4.87e-10 & 1.31e-10 & 4.87e-10 & 5.68e-10 \\
50\(\times\)50 & 3.78e-04 & 3.47e-09 & 3.13e-09 & Timeout & 3.36e-09 \\
\end{tabular}
}
\caption{Precision results for five kernel implementations. The first four differ in the data types used for the algorithm's variables, while the last one corresponds to the Kahan \cite{kahan} summation algorithm. Test cases that exceed 50000 seconds are recorded as Timeout.}\label{tab:prec}
\end{table}

As shown in Table~\ref{tab:prec}, storing both the inner-product and thread-partial result variables in double precision ({\bf{DD}}) can lead to significant deviations from the exact permanent. 
Keeping these variables in quad precision ({\bf{QQ}}), however, increases the execution time by a factor of approximately $4.3\times$. 
Considering that the thread-partial result is updated with $\frac{1}{n}$ frequency compared to the inner-product variable, a viable approach to balance performance and accuracy is to use double precision for the thread-partial variable and quad precision for the inner-product variable, as reported in the second and third columns of Table~\ref{tab:prec}. 
The difference between the techniques in these columns is that the former uses arithmetic implementations following \cite{quad_article}, while the latter follows \cite{quad_forum}. Although the approach in \cite{quad_forum} employs $8$ more floating-point operations ($10$ vs. $18$) than that of \cite{quad_article}, it only slightly improves the accuracy (and only for large matrices), and naturally, it reduces performance. In contrast, Kahan \cite{kahan} summation significantly reduces the relative error compared to \textbf{DD} while incurring only about a 4\% performance penalty. Therefore, in \acro, we leave the choice of using between \textbf{DD} and Kahan \cite{kahan} to the user.

\revise{Although it can be slow, for precision-sensitive applications, we integrated arbitrary precision arithmetic into all the CPU setups in \acro. Users can manually change the number of mantissa bits when precision is more important than performance. We leave the work of arbitrary precision on GPU kernels as future work.} 

%\newpage
\section{Experimental Results}\label{sec:exp}
\acro is written entirely in {\tt C++}/{\tt CUDA} and contains approximately 16,000 lines of code\footnote{{\url{https://github.com/SU-HPC/superman}}}. We conducted our experiments on \revise{three servers, \textbf{Arch-1}, \textbf{Arch-2} and  \textbf{Arch-3}} with \acro compiled using {\tt gcc} 10.5 and {\tt CUDA} 12.3. \textbf{Arch-1} is a local server equipped with two 22-core Intel Xeon Gold 6152 CPUs, 44 cores in total, clocking @2.10GHz and 1 TB of RAM on the host side; and with an Nvidia Quadro GV100 and 32 GBs of memory on the device side. 
The performance tests of the algorithms included in Table~\ref{tab:performance}, as well as the precision tests in Table~\ref{tab:prec}, are conducted using \textbf{Arch-1}. \revise{\textbf{Arch-2} and \textbf{Arch-3} are clusters at TUBITAK ULAKBIM High Performance and Grid Computing Center. \textbf{Arch-2} is used to test \acro's scalability on multiple nodes/GPUs. It has 9 nodes and each node has 8 Nvidia A100 GPUs with 80 GBs of memory. For all the scalability experiments, we used one GPU per node. Last, \textbf{Arch-3} has 48 nodes and 192 Nvidia H200 GPUs in total dedicated to computing the permanent of a $62 \times 62$ matrix in 1.63 days. In addition, the results of the sparse matrix experiments presented in Table~\ref{tab:sparse_experiments} are obtained on this machine.} 

\subsection{Performance of \acro on Dense Matrices}

Table~\ref{tab:performance} compares the GPU‐based permanent computation kernels
for 
dense, randomly generated, double-precision matrices with dimensions $30 \times 30$ to $48 \times 48$. The first row of the table shows the execution time of the multithreaded CPU baseline from~\cite{LUNDOW2022110990}. The CPU code is compiled and executed on {\bf Arch-1} with 44 cores. The GPU kernels are executed on the Quadro GV100 GPU within the same server. 

A kernel $\xb_{mem_1}$--$\Ab_{mem_2}$ is configured to store the row-sum array $\xb$ and the matrix $\Ab$ on $mem_1$ and $mem_2$, respectively, where the storage options are global memory ($glb$), shared memory~($shr$), and registers~($reg$). In the table, each outer block of four rows corresponds to a different kernel configuration, whereas the inner rows present launch parameters, \#blocks and \#threads/block, suggested by the {\tt {CUDA API}} and execution times in seconds. Each kernel has two timings; the first, ``Time", is obtained by using a chunk size $\lceil{\frac{2^{n-1} - 1}{\tau}}\rceil$ as in line~\ref{alg:chunksize} of Algorithm~\ref{alg:parryser} whereas the second, ``Time-CEG", is obtained by using the load distribution strategy described in Subsection~\ref{sec:algo:ceg} which {\bf{C}}oalesces the memory accesses for kernels $\xb_*$--$\Ab_{glb}$ and {\bf{E}}liminates bank conflicts for the kernels $\xb_*$--$\Ab_{shr}$ based on the properties of {\bf{G}}ray codes. Lastly, the last kernel configuration $\xb_{reg}$--$\Ab_{shr}$--$rebuild$ employs the optimization described in Subsection~\ref{sec:algo:rebuild} and uses a different executable for each matrix. The other four kernel configurations use $n_{max} = 63$.

\begin{table}[ht]
\small
\centering
\caption{Comparison of the GPU kernels and the CPU baseline for randomly generated dense matrices with 64-bit entries.
}
\label{tab:performance}
\scalebox{0.8}{
\begin{tabular}{@{}c|l|rrrrrr@{}}
\textbf{Algorithm} & \textbf{Property} 
   & \textbf{30$\times$30} & \textbf{32$\times$32}
   & \textbf{35$\times$35} & \textbf{40$\times$40}
   & \textbf{45$\times$45} & \textbf{48$\times$48} \\
\hline
Lundow and & \multirow{2}{*}{Time} & \multirow{2}{*}{0.65} & \multirow{2}{*}{2.74} & \multirow{2}{*}{23.15} & \multirow{2}{*}{835.50} & \multirow{2}{*}{30323.30} & \multirow{2}{*}{256779.81} \\ 
Markström~\cite{LUNDOW2022110990} & & &  & & &  &  \\ 
\hline
\multirow{4}{*}{$\xb_{shr}$--$\Ab_{shr}$} &  Time (sec)        & 0.91 & 1.16 & 2.16  & 65.95  & 2921.66  & 26164.80 \\ 
& Time--CEG (sec) & 0.87 & 0.99 & 2.08 & 54.59 & 2833.61 & 23121.00 \\
&  \#blocks  & 160  & 160  & 160   & 160    & 160      & 160      \\ 
&  \#threads/block & 160  & 160  & 128   & 96     & 64       & 64       \\ 
\hline
\multirow{4}{*}{$\xb_{shr}$--$\Ab_{glb}$} & Time (sec)     & 0.88 & 1.08 & 2.28  & 55.44  & 1922.73  & 19110.70 \\ 
& Time--CEG (sec)  & 0.89 & 0.98 & 2.16 & 48.69 & 1889.75 & 15507.40 \\
& \#blocks & 160  & 160  & 160   & 240    & 160      & 160      \\ 
& \#threads/block  & 192  & 192  & 160   & 96     & 128      & 128      \\ 
\hline
\multirow{4}{*}{$\xb_{reg}$--$\Ab_{glb}$} &  Time (sec)        & 0.88 & 1.08 & 1.94  & 40.62  & 1052.50  & 12929.30 \\ 
& Time--CEG (sec) & 0.91 & 1.00 & 1.85 & 29.73 & 1016.17 & 8214.59 \\
&  \#blocks  & 80   & 80   & 80    & 80     & 80       & 80       \\ 
&  \#threads/block & 256  & 256  & 256   & 256    & 256      & 256      \\ 
\hline
\multirow{4}{*}{$\xb_{reg}$--$\Ab_{shr}$} &  Time (sec)        & 0.87 & 1.00 & 1.68  & 28.01  & 850.37   & 11224.10 \\ 
& Time--CEG (sec) & 0.86 & 0.94 & 1.67 & 24.41 & 830.76 & 6476.94 \\
&  \#blocks  & 80   & 80   & 80    & 80     & 80       & 80       \\ 
&  \#threads/block & 256  & 256  & 256   & 256    & 256      & 256      \\ 
\hline
\multirow{4}{*}{$\xb_{reg}$--$\Ab_{shr}$--$rebuild$} &  Time (sec)        & 0.87 & 0.95 & 1.29  & 18.31  & 537.67   & 8565.56 \\ 
& Time--CEG (sec) & 0.83 & 0.93 & 1.20 & 14.17 & 528.98 & 4326.23 \\
& \#blocks  & 80   & 80   & 80    & 80     & 80       & 80       \\ 
& \#threads/block & 384  & 384  & 384   & 256    & 256      & 256      \\ 
\end{tabular}
}
\end{table}
\FloatBarrier

The following insights can be obtained for Table~\ref{tab:performance}:
\begin{itemize}[leftmargin=*]

\item Register memory is the most suitable place for $\xb$. Using it mitigates the occupancy problems due to shared-memory limits; the threads/block value of the $\xb_{reg}$ kernels are always larger than those of $\xb_{shr}$ kernels. Furthermore, for $\xb_{shr}$ kernels, the number of threads that can simultaneously run on a single SM decreases when $n$ increases. This is visible in \#threads/block, e.g.,  $\xb_{shr}$--$\Ab_{shr}$ uses 160 threads/block for $n = 30$ and 64 for $n = 45$. 

\item Rebuilding the code is useful for increasing the GPU occupancy which is limited by the number of registers for the $\xb_{reg}$ kernels. In \acro, the fastest configuration is $\xb_{reg}$--$\Ab_{shr}$--$rebuild$. Compared to $\xb_{reg}$--$\Ab_{shr}$, rebuilding yields $53\%$, $59\%$, and $31\%$ improvement on the execution time for $n = 40, 45, 48$ without CEG. With CEG, the improvements due to rebuilding, respectively, are $72\%$, $57\%$, and $49\%$.

\item CEG, the almost conflict-free approach, improves the execution time of $\xb_{reg}$--$\Ab_{shr}$--$rebuild$ by $23\%$, $2\%$, and $50\%$ for $n = 40, 45, 48$, respectively. Since ${\tt gcd}(40, 32) = {\tt gcd}(48, 32) = 8$, the improvements are significant. When $n = 45$, the kernel does not suffer from bank conflicts even without CEG since gcd($n$, 32) $= 1$. As expected, CEG is also useful for $\Ab_{glb}$ kernels since it reduces the number of uncoalesced memory accesses to the matrix.

\item Using GPUs provides significant speedups over the CPU baseline; the fastest configuration $\xb_{reg}$--$\Ab_{shr}$--$rebuild$ obtains $45.6\times$, $56.4\times$, and $30.0\times$ speedups for $n = 40, 45, 48$, respectively. Furthermore, when CEG is used, the speedups increase to $59.0\times$, $57.3\times$, and $59.4\times$. 
\end{itemize}

\subsection{Performance of \acro on Sparse Matrices}

\revise{We examined the square matrices of size between 49--80 from SuiteSparse Matrix Collection\footnote{https://sparse.tamu.edu/} and chose one from each group and collected a set of 18 matrices from 18 different groups. Among these, three matrices have not been impacted by any of our preprocessing algorithms; we discarded those. Among the remaining 15, four matrices are found to be rank-deficient by our preprocessing stages. For the remaining 11 matrices, Table~\ref{tab:sparse_experiments} presents the performance results of \acro with CEG.}

\begin{table}[htbp]
\centering
\small
\scalebox{0.75}{
\renewcommand{\arraystretch}{0.9}
\begin{tabular}{l|lrrr|r}
Matrix & Preprocessing & $n$ & $nnz$ & \#Submatrices & Execution Time \\
\hline
\multirow{4}{*}{{\tt bcspwr02}}
  & None & 49 & 167 & 1 & 4567.23 \\
  & Dulmage-Mendelsohn & 49 & 167 & 1 & 4567.23 \\
  & Forbert-Marx & 30 & 110 & 1 & 0.40 \\
  & Both & 30 & 110 & 1 & 0.40 \\
\hline
\multirow{4}{*}{{\tt d\_ss}}
  & None & 53 & 144 & 1 & TIMEOUT \\
  & Dulmage-Mendelsohn & 53 & 144 & 1 & TIMEOUT \\
  & Forbert-Marx & 30 & 94 & 1 & 0.92 \\
  & Both & 30 & 94 & 1 & 0.92 \\
\hline
\multirow{4}{*}{{\tt curtis54}}
  & None & 54 & 291 & 1 & TIMEOUT \\
  & Dulmage-Mendelsohn & 54 & 291 & 1 & TIMEOUT \\
  & Forbert-Marx & 30 & 164.59 & 143219 & 1362.92 \\
  & Both & 30 & 164.59 & 143219 & 1362.92 \\
\hline
\multirow{4}{*}{{\tt will57}}
  & None & 57 & 281 & 1 & TIMEOUT \\
  & Dulmage-Mendelsohn & 57 & 281 & 1 & TIMEOUT \\
  & Forbert-Marx & 30 & 150.89 & 6834 & 65.89 \\
  & Both & 30 & 150.89 & 6834 & 65.89 \\
\hline
\multirow{4}{*}{{\tt impcol\_b}}
  & None & 59 & 271 & 1 & TIMEOUT \\
  & Dulmage-Mendelsohn & 3.1 & 11.26 & 19 & 18.16 \\
  & Forbert-Marx & 30 & 174 & 1 & 0.76 \\
  & Both & 2.52 & 10.1 & 19 & 0.42 \\
\hline
\multirow{4}{*}{{\tt dwt\_59}}
  & None & 59 & 267 & 1 & TIMEOUT \\
  & Dulmage-Mendelsohn & 59 & 267 & 1 & TIMEOUT \\
  & Forbert-Marx & 30 & 137.56 & 162192 & 1575.81 \\
  & Both & 30 & 137.56 & 162192 & 1575.81 \\
\hline
\multirow{4}{*}{{\tt bfwb62}}
  & None & 62 & 342 & 1 & TIMEOUT \\
  & Dulmage-Mendelsohn & 31 & 171 & 2 & 1.06 \\
  & Forbert-Marx & - & - & - & Stack Overflow \\
  & Both & 29.66 & 181.44 & 9 & 0.47 \\
\hline
\multirow{4}{*}{{\tt ww\_36\_pmec\_36}}
  & None & 66 & 1194 & 1 & IMPOSSIBLE \\
  & Dulmage-Mendelsohn & 6 & 107.63 & 11 & TIMEOUT \\
  & Forbert-Marx & 30 & 872 & 1 & 0.87 \\
  & Both & 3.63 & 80.18 & 11 & 0.42 \\
\hline
\multirow{4}{*}{{\tt west0067}}
  & None & 67 & 294 & 1 & IMPOSSIBLE \\
  & Dulmage-Mendelsohn & 33.5 & 146.5 & 2 & IMPOSSIBLE \\
  & Forbert-Marx & 30 & 135.25 & 3291509 & TIMEOUT \\
  & Both & 30 & 134.23 & 2533556 & 25093.45 \\
\hline
\multirow{4}{*}{{\tt CAG\_mat72}}
  & None & 72 & 1012 & 1 & IMPOSSIBLE \\
  & Dulmage-Mendelsohn & 12 & 128.5 & 6 & 0.40 \\
  & Forbert-Marx & 32.63 & 479.15 & 596728 & TIMEOUT \\
  & Both & 12 & 128.5 & 6 & 0.40 \\
\hline
\multirow{4}{*}{{\tt steam3}}
  & None & 80 & 314 & 1 & IMPOSSIBLE \\
  & Dulmage-Mendelsohn & 3.8 & 14 & 21 & TIMEOUT \\
  & Forbert-Marx & 30 & 143.04 & 88 & 1.66 \\
  & Both & 23.55 & 111.62 & 90 & 1.64 \\
\hline
\end{tabular}
\renewcommand{\arraystretch}{1}
}
\caption{Sparse matrix experiments with 11 real-life matrices: For each matrix, the columns \(n\), \(nnz\), and the number of submatrices ({\em \#Submatrices}) are reported for four preprocessing configurations: (1) None, (2) fine-grain Dulmage-Mendelsohn decomposition, (3) Forbert-Marx decomposition, and (4) Both. For configurations requiring that many submatrices be evaluated, we report the average \(n\) and \(nnz\) across them. The {\em Execution Time} column shows the wall-clock time (s) to complete the configurations, as well as two possible keywords: (1) {\em IMPOSSIBLE}, indicating that computation requires evaluating at least one submatrix with \(n > 63\), and (2) {\em TIMEOUT}, indicating that the computation exceeded the time limit we set to 30,000 seconds.}
\label{tab:sparse_experiments}
\end{table}

\revise{For each matrix, we compute the permanent (1) with no preprocessing, (2) only with fine-grain Dulmage-Mendelsohn decomposition~(Sec.~{~\ref{sec:DMDecompose}}), (3) only with Forbert-Marx decomposition~(Sec.~\ref{sec:FMDecompose}), and (4) with both decompositions. 
Among the 11 matrices, there are seven matrices with $n \leq 63$, and they can be processed without performing any decomposition. \acro computed the permanent of  {\tt{bcspwr02}} in 4567.23 seconds. For the next smaller matrix, {\tt{d\_ss}} with $n = 53$, it could not compute the permanent in 30000 seconds. However, when both decompositions are applied, \acro computes the permanents under 1575.81 seconds for all seven matrices and under 1 second for four of them. Among these seven, Dulmage-Mendelsohn decomposition is useful only for two matrices, {\tt{impcol\_b}} and {\tt{bfwb62}}. For the rest, the Forbert-Marx decomposition is the step that is shown to be useful.
}

\revise{
Among the larger four matrices with $n \geq 64$, the DM decomposition is necessary for {\tt{west0067}} and {\tt{CAG\_mat72}}. That is when DM is removed, either the maximum dimension is $n \geq 64$, i.e., IMPOSSIBLE, or the computation time exceeds 30000 seconds, i.e., TIMEOUT. For the other two matrices, {\tt{ww\_36\_pmec\_36}} and {\tt{steam3}}, a minor reduction in the execution time is observed with DM. Similarly, the Forbert-Marx decomposition is necessary in three out of four matrices. Only for {\tt{CAG\_mat72}}, \acro can compute the permanent without it.}

\subsection{Scalability of \acro}

\begin{figure}[htbp]
    \centering
    \begin{subfigure}[b]{0.65\linewidth}
        \centering
        \includegraphics[width=\linewidth]{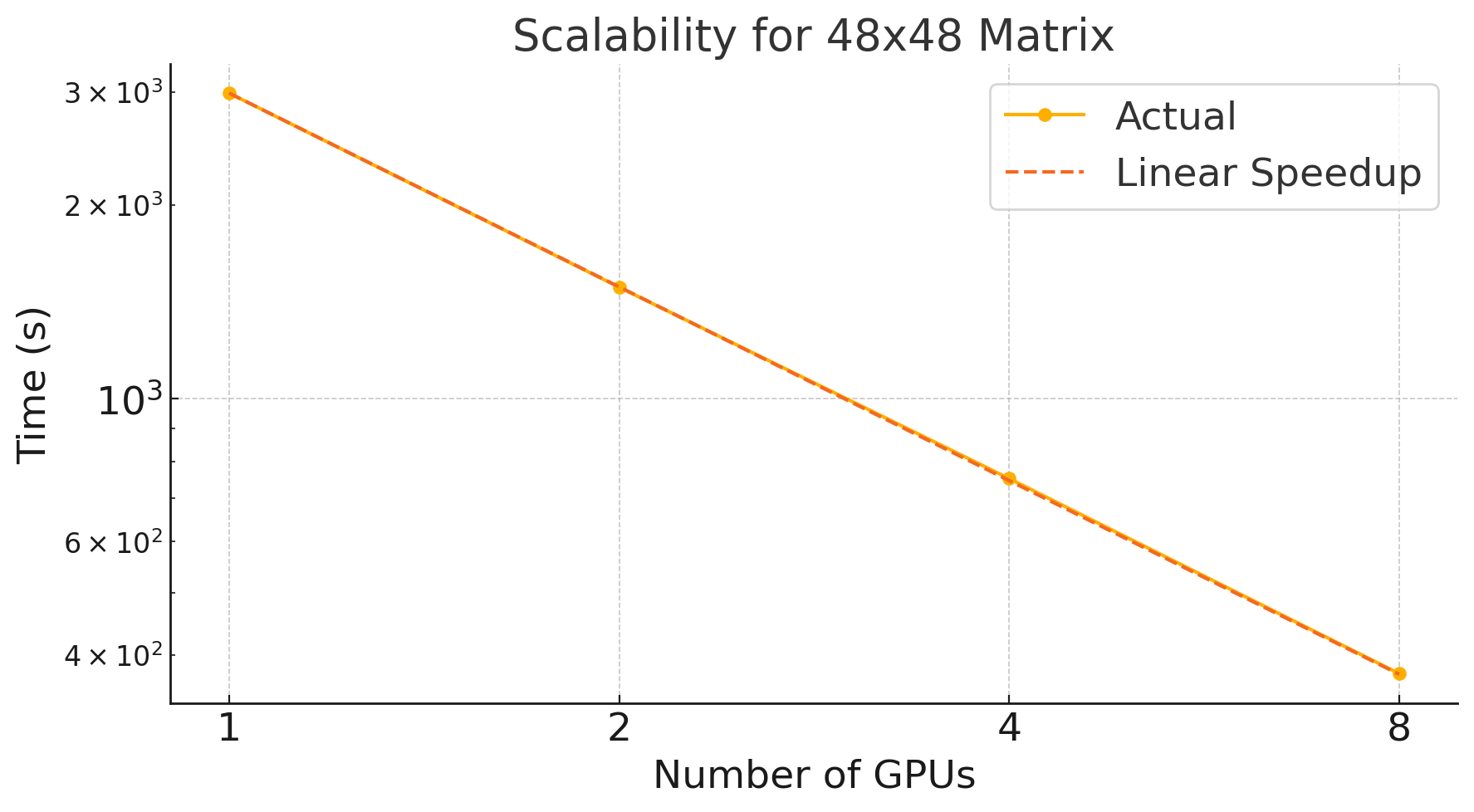}
        \caption{48x48 matrix}
        \label{fig:scalability_48}
    \end{subfigure}
    
    \begin{subfigure}[b]{0.65\linewidth}
        \centering
        \includegraphics[width=\linewidth]{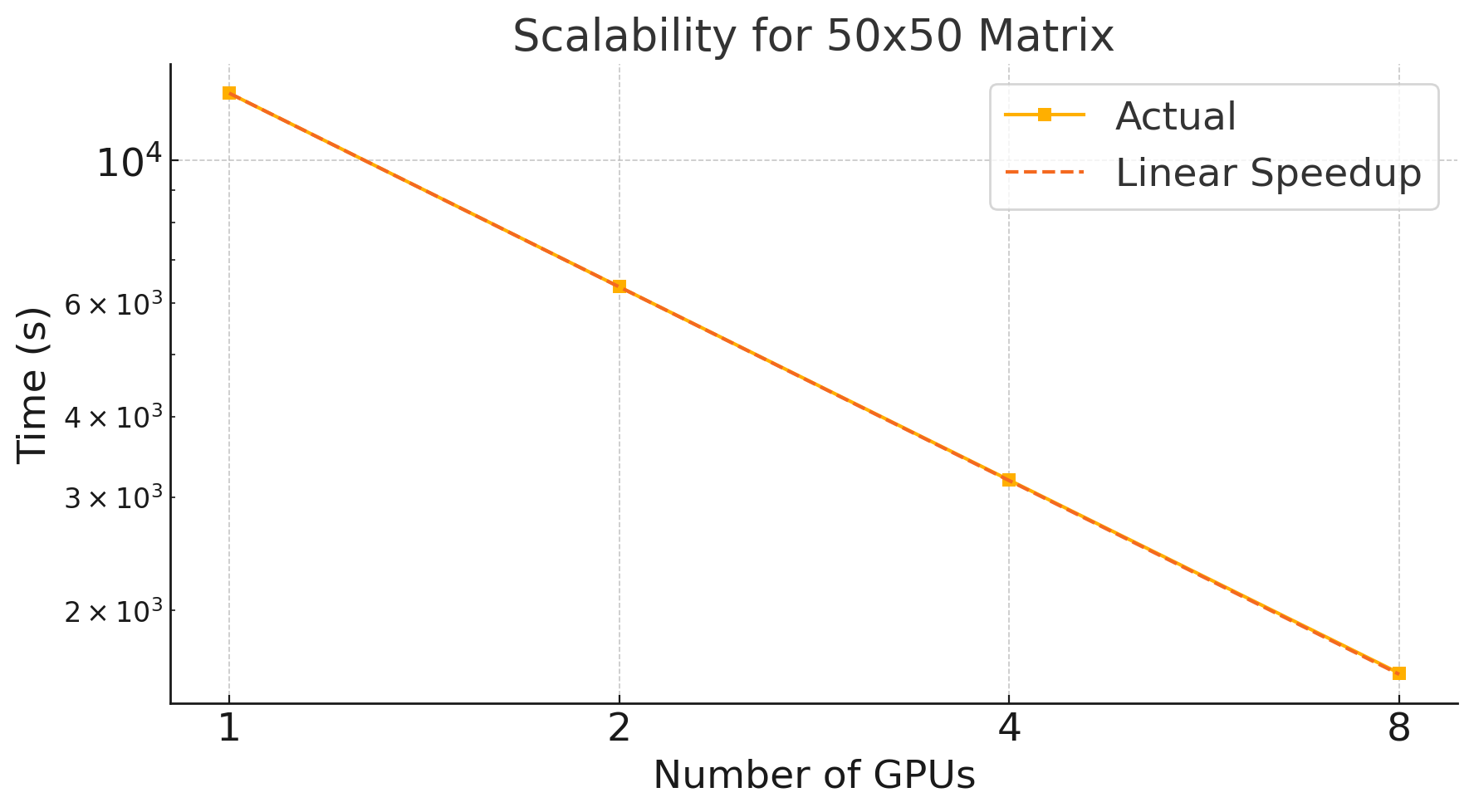}
        \caption{50x50 matrix}
        \label{fig:scalability_50}
    \end{subfigure}
    \caption{Scalability experiments with multiple nodes/GPUs.}
    \label{fig:scalability_experiments}
\end{figure}

\acro supports distributed, multi-GPU permanent computation via Message Passing Interface (MPI). Following the same logic in Algorithm~\ref{alg:parryser}, the iteration space is first divided into the number of GPUs in all the compute nodes as the first-level load distribution. Then each chunk is distributed to the GPU threads by using the CEG optimization described in Subsection~\ref{sec:algo:ceg}. Figure~\ref{fig:scalability_experiments} shows the performance of \acro on $48 \times 48$ and $50 \times 50$ dense matrices with 1, 2, 4, and 8 compute nodes each having a single Nvidia A100 GPU. Since the communication is negligible and only performed to reduce the partial results at the end, the speedups are close to linear with respect to the number of nodes. 

\revise{To further assess the scalability of \acro, we used {\tt Arch-3} with 192 Nvidia H200 GPUs. We established a new record by computing a $62 \times 62$ dense matrix in around 40 hours. This computation requires $294\times$ more computations than the previous $54 \times 54$ problem reported to be solved in the literature. All the entries in the matrix are set to $0.91$ and the permanent error is found to be $10^{-6}$ with Kahan summation.}

\section{Conclusion and Future Work}\label{sec:conc}
We have introduced \acro, a complete tool for the exact computation of matrix permanents on modern GPU architectures. 
\acro supports dense, sparse, real, complex, and binary matrices---a set which covers all matrix types in typical applications.
By exploiting the properties of Gray code–based iterations, carefully placing the data, and distributing the iteration space to the threads, our approach effectively mitigates memory access inefficiencies and bank conflicts. 
\acro also incorporates preprocessing techniques such as DM decomposition. The proposed techniques based on architectural details and careful use of GPUs contribute to reducing the overall computational time with respect to the state of the art.
The experimental results demonstrate \acro's speedups over CPU-based implementations: \acro can be $59\times$ and $86\times$ faster on Nvidia Quadro GV100 and Nvidia A100, respectively, compared to a 44-core CPU baseline. 

\revise{By exploiting the multi-node/multi-GPU configurations, \acro can successfully compute the permanent of a $62 \times 62$ matrix in 1.63 days on 192 Nvidia H200s.}
Our incorporation of advanced precision-enhancing techniques—such as the use of quad precision for critical computations and compensated summation—ensures that numerical accuracy is maintained even as the number of operations grows exponentially with matrix size.

Extending \acro to support hybrid precision and approximation can enable its application to larger matrices and real-life problems. 
Finally, we aim to integrate \acro with wrappers to other ecosystems such as {\tt Python}, thereby increasing its accessibility to researchers across fields such as quantum computing, statistical physics, combinatorics, and graph theory.

\section*{Acknowledgements}
The numerical calculations reported in this paper were mostly performed at TUBITAK ULAKBIM, High Performance and Grid Computing Center (TRUBA resources).

\bibliographystyle{elsarticle-num} 
\bibliography{permanent.bib}
\end{document}